\documentclass[10pt]{IEEEtran}
\usepackage{graphicx}
\usepackage{amssymb}

\usepackage[center,nooneline]{subfigure}
\usepackage{cite}
\usepackage{color}
\usepackage[]{hyperref}

\usepackage{cite}
\usepackage{amsmath,amssymb}

\def\be#1\ee{\begin{equation}#1\end{equation}}
\def\bea#1\eea{\begin{align}#1\end{align}}
\def\bse#1\ese{\begin{subequations}#1\end{subequations}}

\def\1#1{{\hat{{\boldsymbol{#1}}}}}                                 		
\def\2#1{\hat{#1}}                                              		   		
\def\3#1{{\mathbf{#1}}}                                             	   		
\def\4#1{{\boldsymbol{#1}}}                                            		
\def\5#1{{\mathcal#1}}                                                            		
\def\6#1{\bar{#1}}                                                           		
\def\7#1{{{#1}}}                                  		
\def\8#1{\widetilde{#1}}
\def\9#1{\check{#1}}
\def\+#1{{\overset{{\scriptscriptstyle +}}{#1}{}}}                  		
\def\b+#1{{\overset{{\scriptscriptstyle +}}{\mathbf{#1}}{}}}        	
\def\g+#1{{\overset{{\scriptscriptstyle +}}{\boldsymbol{#1}}{}}}    

\def\Eqref#1{Eq.~\eqref{#1}}

\definecolor{dark-green}{rgb}{0.278,0.7,0.4}                    

\definecolor{my-brown}{rgb}{0.69,0.247,0.13}                    

\definecolor{my-purple}{rgb}{0.47,0.12,0.46}                    

\definecolor{my-greenblue}{rgb}{0.129,0.313,0.419}              

\definecolor{my-orange}{rgb}{1,0.5,0.25}                        

\definecolor{my-red}{rgb}{0.745,0,0.2117}                       

\definecolor{my-gray}{rgb}{0.5,0.5,0.5}                         

\definecolor{my-dark-blue}{rgb}{0.1,0.1,0.7}                    

\definecolor{my-indigo}{rgb}{0.29,0.0,0.51}                    

\usepackage{pdfpages}

\begin{document}

\title{Soft Temporal Switching of TL Parameters: Wave-field, Energy Balance, Applications}

\author{Yakir Hadad, \IEEEmembership{Senior Member, IEEE}, Amir Shlivinski, \IEEEmembership{Senior Member, IEEE}
\thanks{Y.Hadad  acknowledges support by the Alon Fellowship 2018-2020.}
\thanks{Y. Hadad is with the School of Electrical Engineering, Tel-Aviv University, Ramat-Aviv, Tel-Aviv, Israel 69978 (e-mail: hadady@eng.tau.ac.il). }

\thanks{A. Shlivinski acknowledges a partial support by the Israel Science Foundation (grant No. 545/15).}
\thanks{A. Shlivinski is  with the Department of Electrical and Computer Engineering, Ben-Gurion University of the Negev, Beer Sheva, Israel, 84105 (e-mail: amirshli@bgu.ac.il).}}

\maketitle

\begin{abstract}
Time-varying guiding structures introduce an additional degree of freedom, besides spatial-variation, that enables better control over the guided wave in a device. Periodically time-modulated structures which are usually considered enable wave control over narrowband signals. However, for ultrawideband short-pulse signals, time-variation in the form of temporal discontinuities is required. Such a setup has recently been proposed as a mean to overcome the Bode-Fano bound on impedance matching. While hard (abrupt) temporal discontinuities are relatively simple to analyze by employing continuity of magnetic flux and electric charge, soft (gradual) temporal switching of the guiding structure parameters is more challenging. This work explores the case of a short-pulse dynamics in a one-dimensional, metamaterial TL, medium with general  smooth time-variation of its parameters. In this time-varying TL, wave-field solutions are obtained by a WKB approach which is more common in the context of gradual spatial variations. Using this methodology a leading order transmitted and reflected waves due to the time-variation  are derived,  followed by a discussion of the energy balance in such switched media. A canonical example of capacitor discharge into a long time-varied TL is given. These results may be used as  analysis/synthesis tools for time-varying wave devices in electromagnetics and acoustics.
\end{abstract}


\section{Introduction}
\label{sec.Introduction}
Wave engineering is the art of tailoring wave phenomena to achieve various functionalities, such as information transmission, signal processing, power absorption,  radiation, imaging, detection, etc.
Conventional wave engineering is based on wave manipulation via \emph{spatial} means\cite{Collin,Kinsler}.
Namely, the desired wave phenomena are obtained and tuned using structural properties of the guiding media to enforce the required bulk wave equations and boundary conditions.
Two typical building blocks are \emph{spatial discontinuities} and \emph{spatial periodicity or modulation} of the guiding configuration. Discontinuities, such as an iris or a step in a waveguide create reflection, affect transmission, and introduce a region of stored reactive energy, thus emulating the presence of a reactive lumped component.
Periodicity, on the other hand, gives rise to another set of effects such as bandgaps, passbands,  distributed resonances, grating lobes, etc. These \emph{two} building blocks are commonly used in wave-devices.



While the use of spatial means for wave control are at the core of wave engineering, already a long time ago an additional venue for wave manipulation has been proposed. That is, the use of \emph{time-variation} of the guiding media.
This approach has been suggested for various purposes such as for signal processing, parametric amplification, and delaylines \cite{Morgenthaler1958, Weinstein1965,Auld1968, Rezende1969, Felsen1970, Fante1971,Agrawal2014, Budko2009, Cervantes-Gonzalez2009, Kunz1966,Chegnizadeh2018,Hayrapetyan2016,Koutserimpas2018},  energy amplification \cite{Lurie2016, Lurie2017, Tretyakov2018,Fleury2018b, Engheta2018Cleo}, non-reciprocity in non-Hermitian time-Floquet systems \cite{Fleury2018}, inverse prism functionality \cite{OL_Caloz2018}, spatial and temporal control of light spectra \cite{Chamanara2019},
temporal-photonic crystal \cite{Halevi2016} with its real-space moving analogue \cite{Skorobogatiy2016}, unusual electromagnetic modes \cite{PRACaloz2018}, mixer-duplexer antenna system \cite{Caloz2017},  for wave pattern engineering \cite{Fort2016, Mattei2017}, as a means to implement synthetic magnetic field \cite{Fang2012}, as well as to break time-reversal symmetry and achieve magnet-less non-reciprocity \cite{Sounas2013, Fleury2014, Estep2014, Hadad2015, Hadad2016, Taravati2017, Sounas2017, Nagulu2018, Fan2018a, Fan2018b, Fan2018c}. The latter is of a particular interest since it hints on the plausibility  of utilizing time-variation to overcome fundamental bounds that are based  on the assumption of \emph{time-invariance}.
%

The examples above expose some of the benefits that are attainable by time-variation.
However, despite the vast research, up to now the spotlight has been primarily aimed at time-variation in the form of \emph{periodic time-modulation}, and consequently, the goals have been typically centered on  \emph{narrowband, continuous-wave}, applications.
In contrast, one may wounder about possibilities to use time-varying media for \emph{ultra-wideband} and \emph{short-pulse} applications by exploiting wave-dynamics with \emph{temporal hard and soft switching} of the guiding media.
In previous publication \cite{Shlivinski2018} we have suggested that an immediate benefit of this approach is the ability to \emph{overcome fundamental limitations} that are associated with the Bode-Fano  bound \cite{Bode, Fano, Acher2009, Pozar, PaiYen, Fleury2014_PRB, Monticone} on the bandwidth of effective impedance matching, with possible further extensions to the Chu-Harrington limit \cite{Chu, Harrington, Gustafsson2007, Gustafsson2009, Gustafsson2015, Thal2006, Schab2018, Capek2017, Wang2010, Wang2014, Wang2017} on small antennas \cite{Shlivinski1997, Shlivinski1999a, Shlivinski1999b}, and constraints on metastructures \cite{Sjoberg2010, Sjoberg2011}. Motivated by \cite{Shlivinski2018}, we believe that  temporal discontinuities have a great potential to  yield wave functionalities, well beyond the state of the art, thus expending the frontiers of wave engineering in electromagnetics and acoustics.

Nonetheless, temporal discontinuities are not realistic, and moreover, occasionally, in order to have more degrees of freedom we would like to have a \emph{sequence} of temporal discontinuities rather than a single one.  This boils down to the problem of wave dynamics in media with softltly varying parameters which is the focus of the present paper. Specifically we consider a one-dimensional guiding medium with simultaneous albeit generally independent variation of the wave velocity and characteristic impedance, or equivalently, permeability and permittivity.  To achieve that goal the paper is organized as follows. Section~\ref{sec.Layout} presents the layout of the problem, concise review on abrupt switching  and a discussion on forward and backward wave propagation in an infinite soft switching TL. Section~\ref{sec.TERMINAL} presents the case of wave propagation due to a terminal voltage at a finite/semi-infinite TL undergoing temporal switching as a generalization of the infinite case. An example of capacitor discharging into a softly switching TL is given in Sec.~\ref{sec.EXAMPLE} with some additional details in the Appendix section. Discussion and conclusions are given in Sec.~\ref{sec.DISCON}.

\section{Layout and preliminaries}
\label{sec.Layout}
We explore wave dynamics in meta-material transmission line with time-varying characteristics, impedance and wave velocity, or equivalently, per-unit-length inductance and capacitance. Such a transmission line may be emulated as a circuit ladder composed of an infinite arrangement of periodic unit cells each consisting a series inductor $L(t)$ and parallel capacitor $C(t)$.
The transmission-line is assumed to be aligned along the $z$-axis.
We denote by $(L_0,C_0)$  [$(L_f,C_f)$] the per-unit-length TL inductance and capacitance at some initial [final] time $t_0$ [$t_f$], where $t_f-t_0\ge T_s$.  Thus, $T_s$ is the switching time between the initial and final times.
Equivalently, during the switching time the characteristic properties of the TL impedance $Z_c(t)=\sqrt{L(t)/C(t)}$ and wave velocity $v(t)=1/\sqrt{L(t) C(t)}$ change between  $\left (Z_0,v_0 \right ) \to \left (Z_f,v_f \right )$.

In a previous publication \cite{Shlivinski2018} we showed that temporally switched TLs may be used to overcome the Bode-Fano criterion for impedance matching. In that work  we have developed an analytical model for the wave dynamics in the idealized case of abrupt switching with $T_s \to 0$ at time $t_s$, namely, with
$\left (Z_c(t_s^-),v(t_s^-) \right )=\left (Z_0,v_0 \right )$ and $\left (Z_c(t_s^+),v(t_s^+) \right )=\left (Z_f,v_f \right )$.
Assuming that initially at time $t < t_s$ there is a pulsed voltage wave with waveform $V_i^+(t)$ propagating in the forward $z$ direction. Continuity of the magnetic flux and electric charge in the TL structure renders, that upon switching at $t_s$, $V_i^+(t)$ is split into two waveforms, one corresponding to a forward propagation wave $V_f^+(t)$ and the other to a backward propagation wave field, $V_f^-(t)$,
\bse
\label{eq.abrupt}
\be
\begin{split}
V_f^+(z,t)&=T V_i^+\left ( \frac{v_f}{v_0}(t-t_s)-\frac{z-z_s}{v_0}\right ),
\\[1ex]
\qquad V_f^-(z,t)&=\Gamma V_i^+\left ( -\frac{v_f}{v_0}(t-t_s)-\frac{z-z_s}{v_0}\right ),
\end{split}
\label{eq.abrupt.a}
\ee
where $z_s=v_0t_s$,  $T$ and $\Gamma$ are the transmission and reflection coefficients. These are given by
\be
T=\frac{1}{2}\left ( \frac{v_f}{v_0}\right )\left [ \frac{Z_f}{Z_0}+1\right ], \quad \Gamma=\frac{1}{2}\left ( \frac{v_f}{v_0}\right )\left [ \frac{Z_f}{Z_0}-1\right ].
\label{eq.abrupt.b}
\ee
Under this temporal switching, the energy change of the wave system is given by \cite{Shlivinski2018}
\be
\Delta \5E= \left \{\frac{1}{2} \left ( \frac{v_f}{v_0}\right ) \left[ \frac{Z_f}{Z_0} +\frac{Z_0}{Z_f}\right ]-1 \right \} \5E_i,
\label{eq.abrupt.c}
\ee
where $\5E_i={\|V_i^+(t)\|^2}/{Z_0}$ is the energy of the initial pulse.
\ese
The transmission and reflection  coefficients in Eq.~(\ref{eq.abrupt.b}) are due to temporal discontinuity, i.e., for abrupt switching. However, for soft (smooth) switching extending over a period of time $T_s>0$, these expressions can not be used as is but instead will be utilized for the construction of the forward and backward waveforms within a consecutive sequence of small abrupt switching framework. This process yields to a WKB-type solutions as discussed next.

\section{WKB-type derivation and the Bremmer series}
\label{sec.WKB}
In the following section we derive the expressions for the forward and backward propagating wavefroms upon propagation in a temporally softly switching (time varying) TL. Since the TL characteristics change smoothly  with time over an extended duration, $T_s>0$, the derivation can use WKB-type arguments by, basically, following two main approaches: $(i)$ a Bremmer-type analysis, see, e.g., \cite{Bremmer1951, Fante1971,Auld1968}, or the more rigorous $(ii)$ ``Generalized'' WKB solution for first order linear differential equation of \cite{Keller1962}. Both formulations gives similar results at least for the first order terms. The Bremmer series approach is more appealing due to its transparent connection to the physical properties of wave propagation, i.e., transmission and reflection, hence we shall follow it for the evaluation of both waveforms and the energy change during switching (for the extension of \Eqref{eq.abrupt}).

\subsection{Forward Wave-field derivation}
\label{ssec.WKB.1}
\begin{figure}[!h]
\centering
\includegraphics[height=80mm]{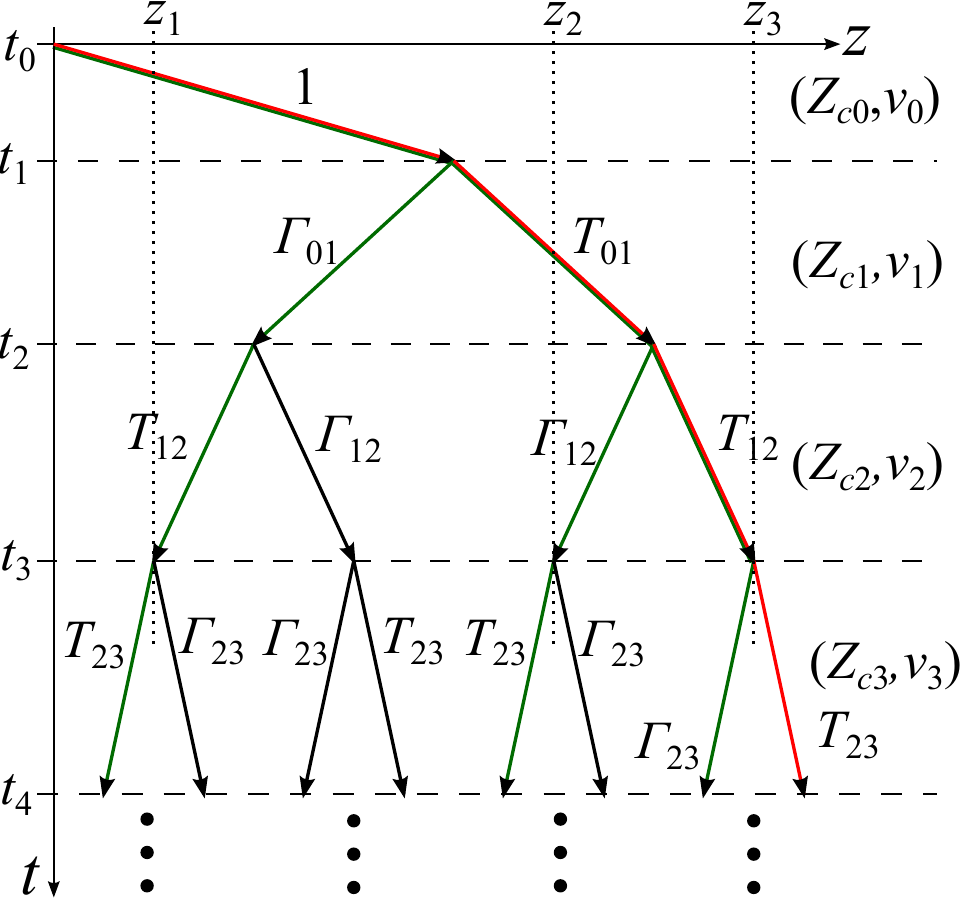}
\caption[]{Wave bouncing diagram in a sequence of successive discontinuities. For small disconttinuities, the reflected wave is order of magnitude smaller than the transmitted wave. Thus, it is possible to aggregate wave contributions according to their order of magnitude. See the red-line for the transmitted wave, and the green-line for the leading order of the reflected waves components.}
\label{Fig.Bounce1}
\end{figure}

We assume that the characteristic properties of the TL $(Z_c(t),v(t))$
softly vary with time. Consequently, a WKB-Bremmer type derivation that is based on ``staircase'' approximation of the TL's characteristic can be applied. In this approximation, it is required that the change of the impedance and wave velocity between adjacent steps is small. To that end, we divide the switching interval into $N$ sub intervals $(t_0,t_1)$, $(t_1,t_2)$, $\ldots$, $(t_{N-1},t_N)$ with corresponding TL's characteristics $(Z_0,v_0)$, $(Z_1,v_1)$, $(Z_2,v_2)$, \ldots $(Z_N,v_N)$. This staircase approximation suggests that at each switch time $t_n$, $n=1 \ldots, t_N$, an abrupt switching takes place with forward and backward wave fields. Thus, in the staircase approximation wave bouncing process occurs.  Figure~\ref{Fig.Bounce1} shows the corresponding binary-tree like bouncing diagram for the this switching process. The weights on each of the branches in Fig.~\ref{Fig.Bounce1} are either the back reflected wave reflection coefficient $\Gamma_{n,n+1}$ or the forward propagating wave transmission coefficient $T_{n,n+1}$, as in  Eq.~\eqref{eq.abrupt.b},
\be
\begin{split}
T_{n,n+1}&=\frac{1}{2}\left ( \frac{v_{n+1}}{v_n}\right )\left [ \frac{Z_{n+1}}{Z_n}+1\right ],
\\[1ex]
\Gamma_{n,n+1}&=\frac{1}{2}\left ( \frac{v_{n+1}}{v_n}\right )\left [ \frac{Z_{n+1}}{Z_n}-1\right ].
\end{split}
\label{eq.switch.b}
\ee

It is easily noted from the bouncing diagram in Fig.~\ref{Fig.Bounce1} that at time $t>t_0$, both the  forward and backward wave-fields are composed of many partial wave contributions that are weighted according to the the amount of transmission or reflection it encountered during switching. Since soft switching is considered, we may approximate $Z_{n+1}=Z_n+\delta_{Z_n}$ with $\delta_{Z_n} \ll Z_n$.  By inserting this approximation  into \Eqref{eq.switch.b} and keeping the lead order terms in $\delta_{Z_n}$ we obtain
\bse
\label{eq.approx}
\bea
T_{n,n+1} &=  \frac{v_{n+1}}{v_n} \sqrt{\frac{Z_{n+1}}{Z_n}} {+\5O\biggl( \left[\frac{\delta_{Z_n}}{Z_n}\right]^2\biggl)},
\label{eq.approx.T}
\\[1ex]
\Gamma_{n,n+1} &= \frac{1}{2} T_{n,n+1} \left [ \frac{\delta_{Z_n}}{Z_n}\right]{+\5O\biggl( \left[\frac{\delta_{Z_n}}{Z_n}\right]^2\biggl)}.
\label{eq.approx.G}
\eea
\ese
Clearly, while the transmission coefficient  through a single infinitesimal temporal discontinuity is ${\cal O}(1)$, the reflection coefficient is much smaller and behaves as ${\cal O}(\delta_{Z_n}/Z_n)$.
Now, consider a scattering process that involves $n$ temporal discontinuities in the time interval $[t_0,t_n]$. In this case,  the overall waveform at time $t_n$ can be decomposed into ($2^n$) waveform components as shown in the bouncing diagram in Fig.~\ref{Fig.Bounce1}. Each of these components can be associated  with a certain order of magnitude that  depends on the number of transmissions and reflections that it experienced during propagation. For example, after $m$ transmissions and $n-m$ reflections the wave component will be ${\cal O}(T^m \Gamma^{n-m})$.
%
In light of that, since the impedance and wave velocity profiles are assumed smooth, the leading order wave component is ${\cal O}(T^n)$ (no reflections), and thus a predominantly forward propagation process takes place.
%
%
This dominant contribution follows the red trajectory marked in Fig.~\ref{Fig.Bounce1}.  This contributions is identified as the WKB approximation of the wave-field, and reads,
\be
\begin{split}
V_f^+&(z,t_{n+1})=\left [\prod_{m=0}^n T_{m,m+1} \right ]\times
\\[1ex]
& V_i^+\!\!\left (\! \frac{v_{n+1}}{v_0}(t_{n+1}\!-\!t_n)\!+\!\sum_{m=0}^n\!\! \frac{v_m}{v_0}(t_{m+1}\!-\!t_m)-\frac{z}{v_0}\!\right)
\label{eq.switch.fwd}
\end{split}
\ee
where in view of Eq.~\eqref{eq.approx.T},
\be
\prod_{m=0}^n T_{m,m+1} \approx \prod_{m=0}^n \frac{v_{m+1}}{v_m} \sqrt{\frac{Z_{m+1}}{Z_m}} = \frac{v_{n+1}}{v_0} \sqrt{\frac{Z_{n+1}}{Z_0}}.
\label{eq.switch.T}
\ee
Approaching the continuum limit where $t_{n+1}-t_n \to 0$, rendering the summation in Eq.~\eqref{eq.switch.fwd} as integration, the final forward wave-field contribution is given by
\be
V_f^+(z,t)=\left [\frac{v(t)}{v_0} \sqrt{\frac{Z(t)}{Z_0}}\right ] V_i^+ \left ( \int_{t_0}^t dt' \, \frac{v(t')}{v_0} -\frac{z}{v_0}\right ).
\label{eq.fwd}
\ee
This forward term is the WKB approximation of the solution.

\subsection{Derivation of the reflected wave}
The first order correction term to the WKB forward wave is a backward propagating wave-field that consists of all wave components of order $\5O(T^{n-1}\Gamma^{1})$. The  wave-bouncing trajectories that correspond to these  components are marked by the green lines in Fig.~\ref{Fig.Bounce1}. Notably, during $n$ discontinuities, there are $n$ wave components of this type that should be summed up. The derivation of this term follows closely the discussion in Eqs.~\eqref{eq.switch.fwd}--\eqref{eq.fwd} albeit somewhat  more involved. Therefore, for the sake of clarity, before presenting the general expression we shall discuss in detail the derivation of the reflection term for the special case of $t_3<t<t_4$ that is depicted in Fig.~\ref{Fig.Bounce1}. From Fig.~\ref{Fig.Bounce1} it follows that at this temporal range, there are three backward propagating contributions that in view of the discussion in Eq.~\eqref{eq.abrupt.a} yield
\bse
\label{eq.back3}
\be
\begin{split}
V_f^-(z,t)=&\Gamma_{01}T_{12}T_{23}V_i^+ \left [ -\frac{v_3}{v_0}(t-t_3)-\frac{z-z_1}{v_0}\right]
\\[1ex]
+&T_{01}\Gamma_{12}T_{23}V_i^+ \left [ -\frac{v_3}{v_0}(t-t_3)-\frac{z-z_2}{v_0}\right]
\\[1ex]
+&T_{01}T_{12}\Gamma_{23}V_i^+ \left [ -\frac{v_3}{v_0}(t-t_3)-\frac{z-z_3}{v_0}\right]
\end{split}
\label{eq.back3.a}
\ee
with $z_1\!=\!v_0(t_1\!-\!t_0)\!-\!v_1(t_2\!-\!t_1)\!-\!v_2(t_3\!-\!t_2)\!=\!2v_0(t_1\!-\!t_0)\!-\!v_0(t_1\!-\!t_0)\!-\!v_1(t_2\!-\!t_1)\!-\!v_2(t_3\!-\!t_2)$, similarly $z_2\!=\!2v_0(t_1\!-\!t_0)\!+2v_1(t_2\!-\!t_1)\!-\!v_0(t_1\!-\!t_0)\!-\!v_1(t_2\!-\!t_1)\!-\!v_2(t_3\!-\!t_2)$ and $z_3\!=\!2v_0(t_1\!-\!t_0)\!+2v_1(t_2\!-\!t_1)\!+\!2v_2(t_3\!-\!t_2)\!-\!v_0(t_1\!-\!t_0)\!-\!v_1(t_2\!-\!t_1)\!-\!v_2(t_3\!-\!t_2)$. Inserting into \Eqref{eq.back3.a} with the approximation in Eq.~\eqref{eq.approx.G} gives
\be
\label{eq.back3.b}
\begin{split}
&V_f^-(z,t)=\frac{1}{2}T_{01}T_{12}T_{23}\left\{\frac{\delta_{Z_0}}{Z_0}V_i^+ \left[ -\tau(t)
+2\frac{v_0}{v_0}(t_1-t_0)\right.\right.
\\[1ex]
&\left.-\frac{z}{v_0}\right]\! + \!\frac{\delta_{Z_1}}{Z_1}\!V_i^+ \!\left[\! -\tau(t)+2\frac{v_0}{v_0}(t_1\!-\!t_0)+2\frac{v_1}{v_0}(t_2\!-\!t_1)-\frac{z}{v_0}\!\right]
\\[1ex]
&+\frac{\delta_{Z_2}}{Z_2}V_i^+ \left[ -\tau(t)+2\frac{v_0}{v_0}(t_1-t_0)+2\frac{v_1}{v_0}(t_2-t_1)\right.
\\[1ex]
&\left.\left.+2\frac{v_2}{v_0}(t_3-t_2)-\frac{z}{v_0}\right]\right\},
\end{split}
\ee
with $\tau(t)=\frac{v_3}{v_0}(t-t_3)+\frac{v_2}{v_0}(t_3-t_2)+\frac{v_1}{v_0}(t_2-t_1)+\frac{v_0}{v_0}(t_1-t_0)$.
\ese
Similar structure follows for later times, thus inductively it yields for $t_n<t<t_{n+1}$
\be
\label{eq.switch.bck}
\begin{split}
&V_f^-(z,t)=\frac{1}{2}\left [\prod_{m=0}^n T_{n,n+1} \right ]  \times
\\[1ex]
&\sum_{m=0}^n \left\{ \frac{\delta_{Z_n}}{Z_n} V_i^+\left[-\frac{v_n}{v_0}(t-t_n)-\sum_{l=0}^{n-1} \frac{v_l}{v_0}(t_{l+1}-t_l)\right. \right.\\[1ex]
&\left.\left. +2 { \sum_{k=0}^{m}} \frac{v_k}{v_0}(t_{k+1}-t_k) -\frac{z}{v_0} \right] \right\}.
\end{split}
\ee
Approaching the continuum limit, we note that $\frac{\delta_{Z_n}}{Z_n} \to \left (\frac{d \ln Z(t)}{dt}\right)dt$, and the final expression for the first order correction term is obtained
\be
\begin{split}
&V_f^-(z,t)=\frac{1}{2} \left [\frac{v(t)}{v_0} \sqrt{\frac{Z(t)}{Z_0}}\right ] \times
\\[1ex]
&\int_{t_0}^t\! \!dt' \frac{d \ln Z(t')}{dt'} V_i^+\!\! \left [\! -\!\! \int_{t_0}^t \!\! dt'' \frac{v(t'')}{v_0}\!+\!2\!\!\int_{t_0}^{t'} \! \!\!\!dt'' \frac{v(t'')}{v_0} \!-\!\frac{z}{v_0}\right]\!.
\end{split}
\label{eq.bck}
\ee
As a correction term to the forward propagating WKB wave-field in Eq.~\eqref{eq.fwd}, this term is identified as the leading order term of the reflected wave-field.

The next correction term involves all wave components of order $\5O(T^{n-2}\Gamma^2)$. This is a forward propagating wave-field that compensates the WKB contribution for the changes in time of $(Z(t),v(t)$. However, since it consists of two reflections (at two different switches) it is negligibly small and will not be considered further. Nevertheless, its derivation follows the same lines as above.

Lastly, we emphasize that while the discussion so far did not make any explicit reference to the field's waveform, implicitly this formulism assumes that the duration of the switching, $T_s$, is much larger than the pulse width.
Moreover, in the aforementioned derivation of the forward wave and the first order reflection we have  followed Bremmer-type arguments that are intimately connected with the wave-field dynamics. An alternative approach that gives similar results is based on an asymptotic solution of linear differential equation to give a generalized type of WKB solution was treated in \cite{Keller1962}. The latter, while easily used for the two first order terms in the approximation of the wave-field, yields less transparent higher order terms.


\subsection{Energy balance derivation}
\label{ssec.WKB.2}
The discussion in the previous section concerned the forward and backward wave dynamic. However the discussion is not complete without the assessment of the energy balance during switching. The energy balance in a single abrupt switching was discussed in \cite{Shlivinski2018} and outlined above in Eq.~\eqref{eq.abrupt.c}. In this section we extend the discussion to the soft switching following  arguments akin to that used in the previous sections for the wave-fields.

The energy balance at time $t_n<t<t_{n+1}$ is obtained by summing up all the partial energy balances during each of the step-like abrupt switchings:
\be
\Delta \5E=\sum_{m=0}^{n-1} \Delta \5E_m,\label{eq.enrg.blnc}
\ee
where {$\Delta \5E_m = \Delta e_{m,m+1} {\|V_m^+(t)\|^2}/{Z_m}$} denotes the energy change during switching at $t=t_{m+1}$. We use the  energy balance expression in Eq.~\eqref{eq.abrupt.c}  for a  single abrupt switching in order to write
\be
\Delta e_{n,n+1}\!=\! \left \{\frac{1}{2} \!\left ( \frac{v_{n+1}}{v_n}\right )\! \!\left[ \frac{Z_{n+1}}{Z_n}\! +\!\frac{Z_n}{Z_n+1{}}\right ]\!-\!1\! \right \} \approx \frac{\delta v_n}{v_n}
\label{eq.De}
\ee
where the first order approximation on the right hand side applies for soft switching with $Z_{n+1}=Z_n+\delta_{Z_n}$ and $v_{n+1}=v_n+\delta_{v_n}$ where $\delta_{Z_n} \ll Z_n$ and $\delta_{v_n} \ll v_n$. Interestingly, with regard to the characteristic impedance, this expression is variational which means that small  variation in the impedance during switching appears as second order corrections.
\bse
\label{eq.blnc.stp}
Looking at the limited time scenario depicted in Fig.~\ref{Fig.Bounce1}, it can be seen that at the first switching the balance is given by Eq.~\eqref{eq.abrupt.c} where with it yields
\be
\label{eq.blnc.stp.a}
\Delta \5E_0= \Delta e_{01}\frac{\|V_i^+(t)\|^2}{Z_0} \approx \frac{\delta v_0}{v_0} \frac{\|V_i^+(t)\|^2}{Z_0}.
\ee
For the second switching at $t=t_2$, there are two contributions that gives the first order approximation:
\be
\label{eq.blnc.stp.b}
\begin{split}
&\Delta \5E_1=\\[1ex]
&\Delta e_{12}\frac{\left \|T_{01}V_i^+\left(\frac{v_1}{v_0}t\right)\right \|^2}{Z_1} \!+\! \Delta e_{12}\frac{\left \|\Gamma_{01}V_i^+\left(-\frac{v_1}{v_0}t\right)\right \|^2}{Z_1}
\\[1ex]
&= \Delta e_{12} \left [ T_{01}^2+\Gamma_{01}^2\right ] \frac{v_0}{v_1} \frac{Z_0}{Z_1} \frac{\|V_i^+(t)\|^2}{Z_0} \\[1ex]
&=\frac{1}{2} \Delta e_{12} \frac{v_1}{v_0} \left [ \frac{Z_{1}}{Z_0}+\frac{Z_{0}}{Z_1}\right ] \frac{\|V_i^+(t)\|^2}{Z_0}
\\[1ex]
&\approx \frac{\delta v_1}{v_1} \frac{v_1}{v_0} \frac{\|V_i^+(t)\|^2}{Z_0}= \frac{\delta v_1}{v_0} \frac{\|V_i^+(t)\|^2}{Z_0}
\end{split}
\ee
where we used
\be
T_{n,n+1}^2+\Gamma_{n,n+1}^2 \!=\! \frac{1}{2} \!\left ( \frac{v_{n+1}}{v_n}\right )^2\!\! \left ( \frac{Z_{n+1}}{Z_n}\right ) \!\!\left [ \frac{Z_{n+1}}{Z_n} \!+\! \frac{Z_{n}}{Z_{n+1}}\right ]\!.
\label{eq.T2G2}
\ee
In a similar manner, for the $n$'th switching at $t=t_n$ it follows that
\be
\label{eq.blnc.stp.d}
\Delta \5E_n \approx \frac{\delta v_n}{v_0} \frac{\|V_i^+(t)\|^2}{Z_0}.
\ee
\ese
Then, by summing over all the partial contributions in Eq.~\eqref{eq.enrg.blnc}, and approaching the continuum limit we obtain the following remarkably simple result for the energy change in a {smooth (soft)} variation,
\be
\Delta \5E =\left (\frac{v(t)}{v_0} -1\right ) \frac{\|V_i^+(t)\|^2}{Z_0}
\label{eq.enrg.blnc.2}
\ee
which is similar to  Eq.~\eqref{eq.abrupt.c} for a single abrupt switching \emph{of the velocity only switching}, albeit here we have switched both the velocity and the impedance. The omission of the ``impedance'' term comes due to the first order nature of the approximation and since any small changes in the impedance appears as a second order correction.

\section{Terminal voltage WKB-type derivation}
\label{sec.TERMINAL}
The discussion so far assumed that the TL is infinite,  and moreover that, initially, prior to the  switching time at $t=t_0$, there is a voltage waveform $V_i^+(t)$ propagating in it. However, it would be useful to consider the more realistic scenario in which the TL is finite or semi-infinite and it is excited by a source at its terminal.  Let us set the terminal location at  $z=0$, there,  a voltage $V_0(t)$ and current $I_0(t)$ are imposed by an external source. At the TL terminal the relation $V_0(t)=Z(t) I_0(t)$ is satisfied where, as before, $Z(t)$ is the time dependent TL's characteristic impedance. This section is devoted for the derivation of the forward propagating WKB-type voltage wave. This derivation uses the expression derived in the previous section and in particular of \Eqref{eq.fwd}.
Given a terminal voltage waveform $V_0(t)$ that initiates at $t=0$, it may be expressed as
\begin{equation}
V_0(t)=\int_0^{\infty}V_0(\tau)\delta(t-\tau)d\tau
\end{equation}
where $\delta(t)$ is dirac's delta. This representation implies that the input signal can be expressed as a superposition of weighted impulses, each excited at time $t=\tau$. However, since the TL system is linear (albeit time-variant) we may write the excited wave as a superposition integral of weighted impulse responses $h_f^+(z,t;\tau)$.
For an impulse signal at time $t=\tau$, $V_i^+(t)=\delta(t-\tau)$, using Eq.~(\ref{eq.fwd}), the forward propagating wave at time $t$ and location $z$ on the line, reads,
\begin{equation}
h_f^+(z,t;\tau)=\left[\frac{v(t)}{v(\tau)}\sqrt{\frac{Z(t)}{Z(\tau)}} \right]\delta\left( \int_\tau^t dt' \frac{v(t')}{v_0} -\frac{z}{v_0} \right).
\end{equation}
It is important to emphasize that this solution that was obtained for the infinite line case, is valid, also, here only because the excited pulse width is zero. This is the impulse response of the time-variant line. Once we have it, we can calculate the response for any excitation using superposition,
\be
\begin{split}
  V_f^+(z,t) =&  \int_{0}^{\infty} \!\!d\tau \, V_0(\tau) h_f^+(z,t;\tau)
  \\[1ex]
   =& \frac{v(t)}{v(\tau_0)}\sqrt{\frac{Z(t)}{Z(\tau_0)}}  V_0(\tau_0),
   \end{split}
   \label{eq.Vf.4}
\ee
where $\tau_0$ satisfies the integral equation
\be
\int_{\tau_0}^t dt' \, v(t') -z=0, \qquad \tau_0<t_0.
\label{eq.tau_0}
\ee
The solution of this equation gives the time $\tau_0=\tau_0(z,t)$ at which an infinitesimal wave contribution appears at the terminal of the TL in order to reach $z$ at time $t$ (for the special case of $v(t)=v_0$ it follows that $\tau_0=t-z/v_0$).
It is immediately noted that the expression in \Eqref{eq.Vf.4} is different than the one in \Eqref{eq.fwd}. However this  is obvious upon noting the different scenarios (as mention in the beginning of this section): in \Eqref{eq.fwd} it was assumed the wavefield is wholly contained in the TL during before and during switching while here the wavefield is imposed at the boundary (at $z=0$) during switching. Moreover, in the present case, different sections of the waveform (along the time) undergo different switching, i.e., while the leading edge ``sense'' the entire switching period the trailing edge ``sense'' only parts of it. This has further implications as for example non-uniform different temporal compression{/expansion} and amplification of the waveform. Note that this result generalize previous results in, for example, \cite{Fante1971} that assumed switching of only the electric permittivity. Here, on the other hand, it is assumed the both the impedance and velocity, i.e., generally, the permeability and permittivity, can change independently.

The backward propagating (reflected) wave-field $V_f^-(z,t)$ due to the forward propagating wave $V_f^-(z,t)$ can be calculated in much the same way by starting with \Eqref{eq.bck}. However, special attention should be given to the fact that this wave propagates toward the terminal at $z=0$ and due to impedance mismatch, there, parts of  it back-reflecte as delayed forward waves. Hence, it,  therefore, may happen that all these contributions overlaps in time and space in the vicinity of the terminal. Assuming gradual impedance and velocity changes during switching, renders weak (negligible) $V_f^-$ such that $V_f^+$ gives {an accurate approximation of the propagating wave.}

While the discussion in this and previous sections was focused on finding the wave field solutions in a heuristic and intuitive way using the Bremmer series and tracing wave-fronts as representing fundamental solutions of the wave-equation, one may solve the TL's set of differential equations: $\partial_z Q(z,t) = -C(t) \partial \Phi(z,t)$ and $\partial_z \Phi(z,t) = -L(t) \partial Q(z,t)$ where $Q(z,t)=C(t) V(z,t)$ and $\Phi(z,t)=L(t) I(z,t)$ and with an appropriate boundary conditions, either for the infinite TL case or the finite case (Sec.~\ref{sec.WKB} and Sec.~\ref{sec.TERMINAL}, respectively). A general solution of these equation can follow a spectral transformation of the the $z$ coordinate, rendering them a standard set of first order linear differential equations admitting an exponential type integral solution. Applying successive approximation yields {the corresponding WKB type solution in the spectral domain}, \cite{Keller1962}. Transforming back the spectral solution to the spatial, $z$, coordinate gives similar results for the wavefields to those obtained in Sec.~\ref{sec.WKB} and Sec.~\ref{sec.TERMINAL}. In the special cases where $C(t)$ and $L(t)$ are ``convenient'' functions of time, an exact solution of the wave-equation may be formulate, see, e.g. \cite{Hayrapetyan2016}.

\section{Example}
\label{sec.EXAMPLE}
\begin{figure}[!h]
\centering
\includegraphics[width=85mm]{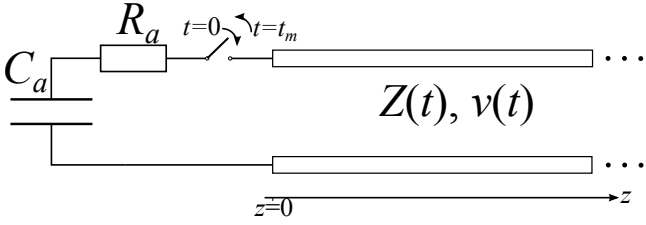}
\caption[]{Capacitor discharge scheme. At time $t=0$ a switched is closed and the capacitor starts to discharge into the TL. The TL characteristics are time-varying in a way to maximize the energy delivery from the capacitor to the line.}
\label{Fig.Discharge}
\end{figure}

In the discussion above we have derived analytical solutions for Ultra-Wide-Band short-pulse propagation in, and excitation of a softly time-varying TL. Here, we show how this formalism can be used in order to solve a simple canonical problem of lumped capacitor discharge into a softly time-varying TL. The analyzed configuration is shown in  Fig.~\ref{Fig.Discharge}.  The TL characteristics varies with time between two states, initial and final. In this process both the impedance and the wave velocity of the line may be changed with time, however, in the WKB approximation, since reflection is negligibly small, the discharged capacitor effectively sees only the time-varying TL characteristic impedance as a load, with no additional correction.  Once the pulse is in the line the velocity time-variation may also affect the propagation dynamics.

Assume that for $t<0$, the capacitor , $C_a$, is charged with $-Q_0$. At $t=0$ a switch, $sw$ is closed and the capacitor starts to discharge through a fixed resistor $R_a$ and a softly varying  TL with characteristic impedance $Z(t)$. The initial TL impedance at $t=0$ is $Z_0$.  In the discharge process some of the capacitor's initial energy is dissipated on the series resistor $R_a$ while the rest is transmitted in the line. Our goal is to optimize the energy delivery into the line by changing the load $Z(t)$ (i.e., changing the TL's characteristic impedance). For that purpose we solve analytically the variational problem that leads to the best characteristic impedance trajectory in time $Z(t)$ (See the Appendix for the optimization details) to give
\be
Z(t)=R_a \zeta(t), \qquad \zeta(t)=\zeta_0-\frac{t}{\tau_a},
\label{eq.exmpl.1}
\ee
where $\zeta_0=Z_0/R_a$ and $\tau_a=R_a C_a$.
This time-varying impedance renders constant discharge current $I_d=Q_0/C_a(R_a+Z_0)$.
As can be seen in \Eqref{eq.exmpl.1}, during discharge $Z(t)$ is a strictly monotonic decreasing function of $t$. Since $Z(t)$ is restricted to be positive  for all $t$, the minimal value of the TL characteristic impedance is set to $Z_m=R_a \zeta_m\ll Z_0$, this takes place at $t_m=(\zeta_0-\zeta_m)\tau_a$. Exactly at this instant, $t=t_m$, the switch, $sw$, is opened again, thus ending the discharge process. The remaining $C_a$ charge is given by $Q_m=- Q_0(1+\zeta_m)/(1+\zeta_0) \ll Q_0$ \footnote{Note that if at time $t=t_m$ the switch remains closed and $Z(t)=Z_m$, the remaining discharge is exponentially decaying to zero at $t \to \infty$}. Since the current $I_d$ flows into the time varying TL with characteristic impedance $Z(t)$ it produces a terminal voltage at $z=0$ of
\be
V_0(t)= \frac{Q_0}{C_a} \frac{{\zeta(t)}}{1+\zeta_0} \bigl [ H(t) -H(t-t_m)\bigl],
\label{eq.exmpl.2}
\ee
which has a trapezoidal waveform where $H(t)$ denotes the Heaviside step function.

The discussion above was about the capacitor's discharge that creates a voltage $V_0(t)$ at the TL's terminal at $z=0$. This voltage, in turn, generates a forward propagating voltage wave in the TL (which may be either semi-infinite or with a matched load) according to \Eqref{eq.Vf.4}. The following discussion aims to explore and demonstrate some of the propagation characteristics.

Let us assume that the time varying TL is implemented as a periodic circuit of series inductors $L(t)$ and parallel capacitors $C(t)$ per-unit-length, such that $Z(t)=\sqrt{{L(t)}/{C(t)}}$ and $v(t)={1}/{\sqrt{L(t) C(t)}}$. Following \Eqref{eq.exmpl.1} the normalized TL's characteristic impedance ($\zeta(t)=Z(t)/R_a$) can be extended along $t$ as
\be
\zeta(t)=\zeta_0 -\frac{t}{\tau_a} H(t) -\left (\zeta_0-\zeta_m-\frac{t}{\tau_a} \right ) H(t-t_m).
\label{eq.exmpl.3}
\ee
As discussed above, the choice of $v(t)$ is detached from that of $Z(t)$ since it affects only the wave propagation and can be done arbitrarily. Nevertheless, there are some few favourable such choices, as follows:
\\[1ex]
$\bullet$ \textbf{Constant velocity scheme:} The pulse width of the terminal voltage, $V_0(t)$ is $t_m$, \Eqref{eq.exmpl.2}, however in view of Eqs.~\eqref{eq.Vf.4} and \eqref{eq.tau_0}, while propagating in a TL with time dependent velocity, $v(t)$, the pulse width generally scales up/down (pulse compression/expansion). In order to maintain the same pulse width, it follows that the velocity should be time independent, $v(t)=v_0$. Thus, following \Eqref{eq.tau_0}, leading to {$\tau_0=t-z/v_0>$ while propagating.}
However, this requirement is redundant since $V_0(\tau_0)=0$ for $\tau<0$. Moreover, since, $Z(t)$ and $v_0$ are set, \Eqref{eq.Vf.4} renders,
\be
\begin{split}
V_f^+(z,t)&=\sqrt{\frac{Z(t)}{Z(t-z/v_0)}} V_0(t-z/v_0)
\\[1ex]
&
= \frac{Q_0}{C_a}\, \frac{\sqrt{\zeta(t)\,\zeta(t-z/v_0)}}{1+\zeta_0}\times
\\[1ex]
&
\bigl [ H(t-z/v_0) -H(t-z/v_0-t_m)\bigl].
\end{split}
\label{eq.exmpl.4}
\ee
Though this wave velocity setup prevents pulse compression, yet, there is an unavoidable multiplicative time-dependent amplitude scaling of the propagating contribution. In order to maintain this ``constant velocity'' setting, a synchronous time variation of both the inductance and capacitance during switching is required (see a more general case below).

\begin{figure*}[!ht]
\centering
\subfigure[$\zeta_0=2$]{%
     \label{Fig.Exmpl2.a}%
      \includegraphics[height=50mm]{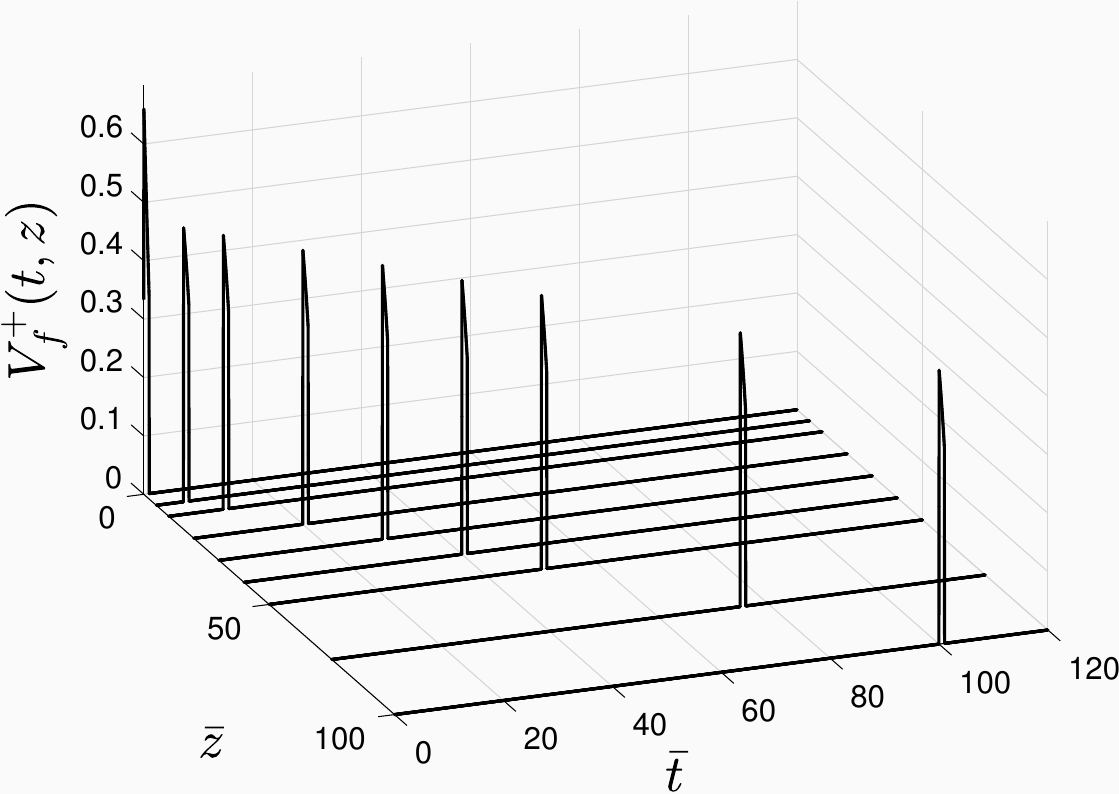}}
\hspace{10mm}
   \subfigure[$\zeta_0=5$]{%
     \label{Fig.Exmpl2.b}%
     \includegraphics[height=50mm]{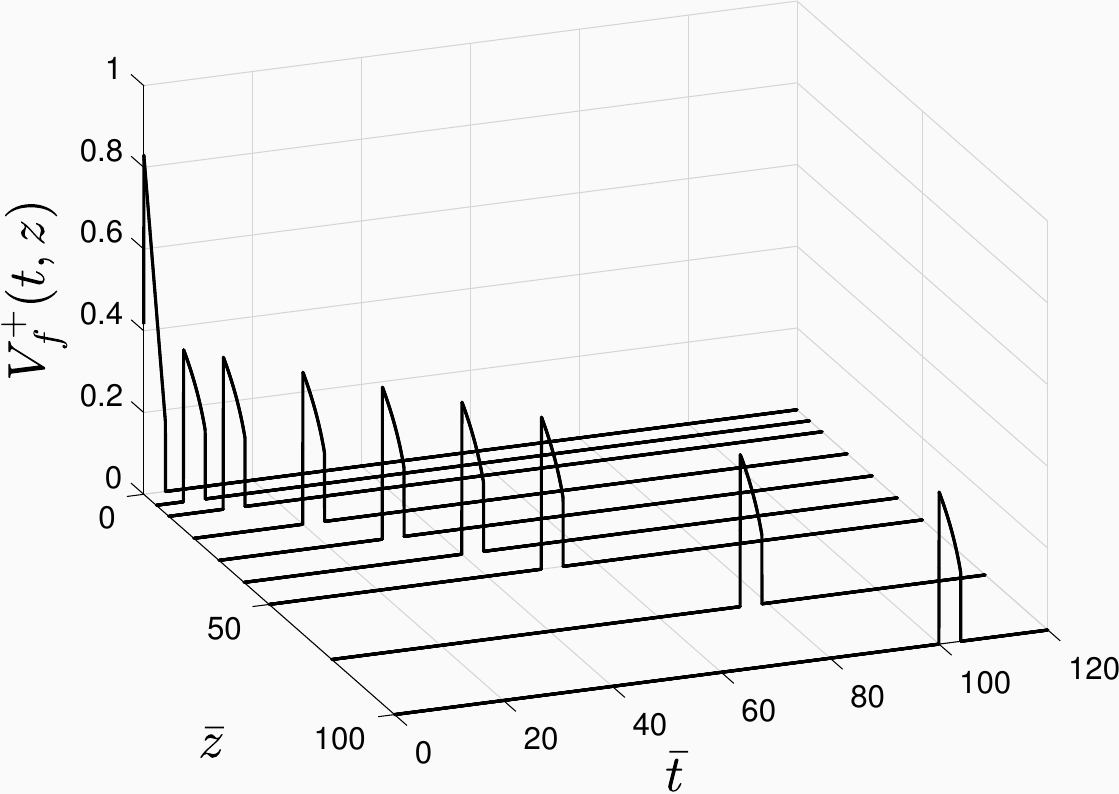}}
\vspace{1ex}
\centering
   \subfigure[$\zeta_0=10$]{%
     \label{Fig.Exmpl2.c}%
     \includegraphics[height=50mm]{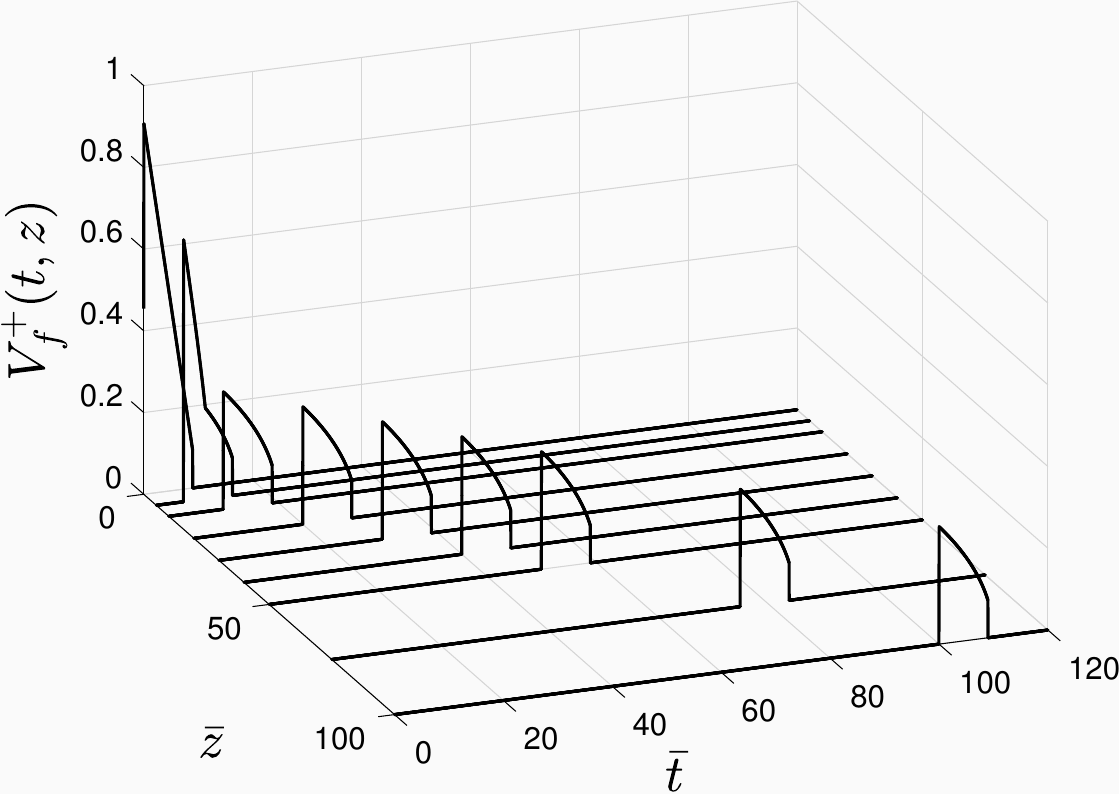}}
\hspace{10mm}
   \subfigure[$\zeta_0=20$]{%
     \label{Fig.Exmpl2.d}%
     \includegraphics[height=50mm]{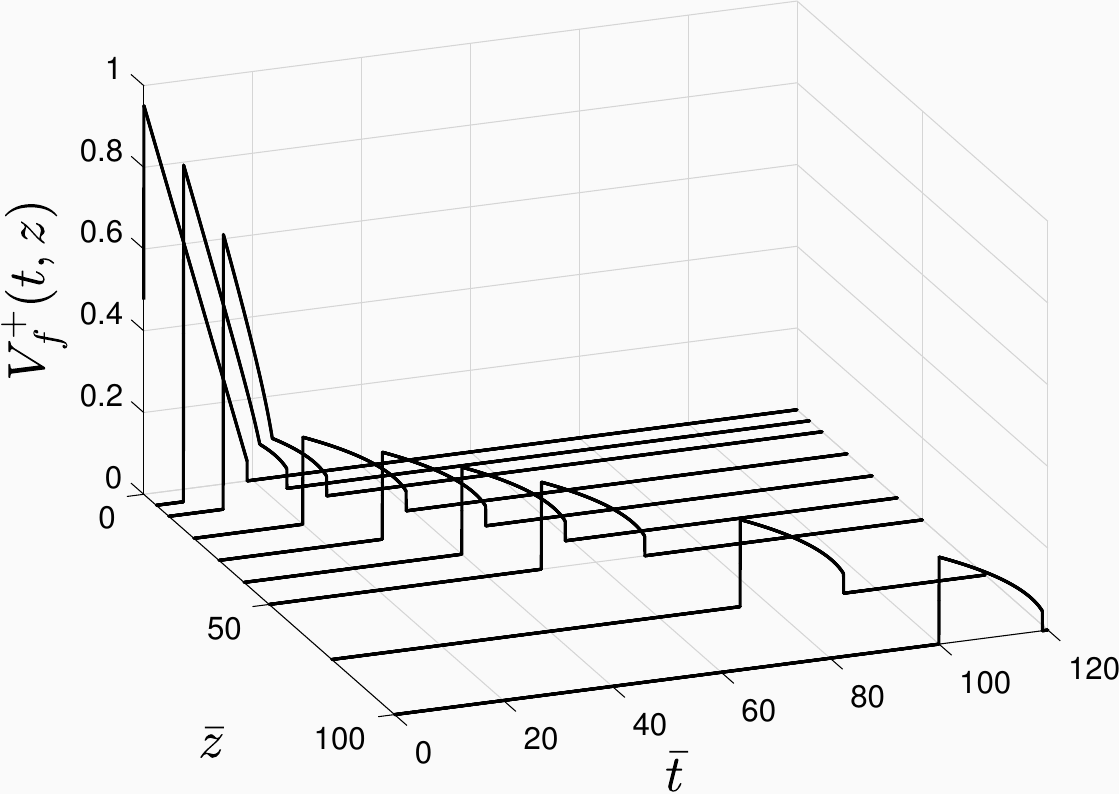}}
\caption[]{Normalized space-time snapshots of the pulse propagation in the a softly time varying TL with constant velocity $v_0$. The axes are normalized such that $\6z=z/l_a$ with $l_a=\tau_a v_0$ and $\6t=t/\tau_a$.}
\label{Fig.Exmpl2}
\end{figure*}

Figure~\ref{Fig.Exmpl2} depicts space-time snapshots of the wave propagation for the capacitor's discharging circuit (Fig.~\ref{Fig.Discharge}) in a constant velocity TL with $Q_0/C_a=1$, $\zeta_m=1$ at $\6z=z/l_a=0,5,10,20,30,40,50,75,100$ (with $l_a=\tau_a v_0$), $\6t=t/\tau_a$ and for several cases of initial characteristic impedances $\zeta_0=2,5,10,20$ in Figs.~\ref{Fig.Exmpl2.a}, \ref{Fig.Exmpl2.b}, \ref{Fig.Exmpl2.c}, \ref{Fig.Exmpl2.d}, respectively. Viewing these figures: $(i)$ the effect of the impedance switching on the waveform is readily noted by comparing the terminal voltage at $z=0$ and the voltages at points $z$ deep inside the TL; $(ii)$ For early arrival times of the leading edge of the wavefront, that are prior to the cessation of the switching (The latter takes place at $\6t_m=t_m/\tau_a=1,4,9,19$, in Figs.~\ref{Fig.Exmpl2.a}, \ref{Fig.Exmpl2.b}, \ref{Fig.Exmpl2.c}, \ref{Fig.Exmpl2.d}, respectively), the behavior of the waveform is different than for later times ($\6t>\6t_m$) due to the interplay between the $\zeta(t)$ and $\zeta(t-z/v_0)$ terms in \Eqref{eq.exmpl.4} that change with relative spatial delay. Moreover, $(iii)$ for $t>t_m$, where $\zeta(t)=\zeta_m$, the wave propagation becomes a shift invariant process.

\vspace{1ex}
\noindent$\bullet$ \textbf{Capacitor only switching scheme:} Though favourable in keeping the same pulse width during propagation, the ``constant velocity'' scheme requires the synchronous change of both TL per-unit-length capacitance and inductance. However, while time-varying capacitance is relatively simple to implement (e.g., using varactor diodes), a continuous time-variation of inductance is somewhat more challenging. Therefore, it is interesting to analyze the practical case of a metamaterial TL with only time-varying capacitance.

To that end, assume that the per-unit-length inductance is maintained constant $L(t)=L_0$. In that case the per-unit-length capacitance is given by $C(t)=L_0/[Z(t)]^2$.  Yielding
\be
v(t)=\frac{Z(t)}{L_0}=\frac{R_a}{L_0} \zeta(t).
\label{eq.exmpl.5}
\ee
It is interesting to note that following \Eqref{eq.exmpl.1}, $v(t)$ is also a monotonic decreasing function of $t$. Consequently, during propagation the wave-field slows down from the initial velocity $\zeta_0 R_a/L_0$ to the final velocity $\zeta_m R_a/L_0$ and, thus, energy is absorbed from the wave system during switching.

Following, \Eqref{eq.tau_0}, $\tau_0$ is obtained by solving
\be
\int_{\tau_0}^t dt' \, \zeta(t') =\frac{L_0}{R_a} z.
\label{eq.exmpl.6}
\ee
Thus yielding the forward propagating wave
\be
\begin{split}
&V_f^+(z,t)=\left[ \frac{Z(t)}{Z(\tau_0)}\right ]^{3/2} V_0(\tau)
\\[1ex]
&
= \frac{Q_0}{C_a}\frac{\bigl [ \zeta(t) \bigl]^{3/2}}{(1+\zeta_0)}  \,\frac{ \bigl [ H(\tau_0(z,t)) -H(\tau_0(z,t)-t_m)\bigl]}{\sqrt{\zeta(\tau_0(z,t))}}.
\end{split}
\label{eq.exmpl.7}
\ee
\begin{figure*}[!ht]
\centering
   \subfigure[$\zeta_0=2$]{%
     \label{Fig.Exmpl1.a}%
     \includegraphics[height=50mm]{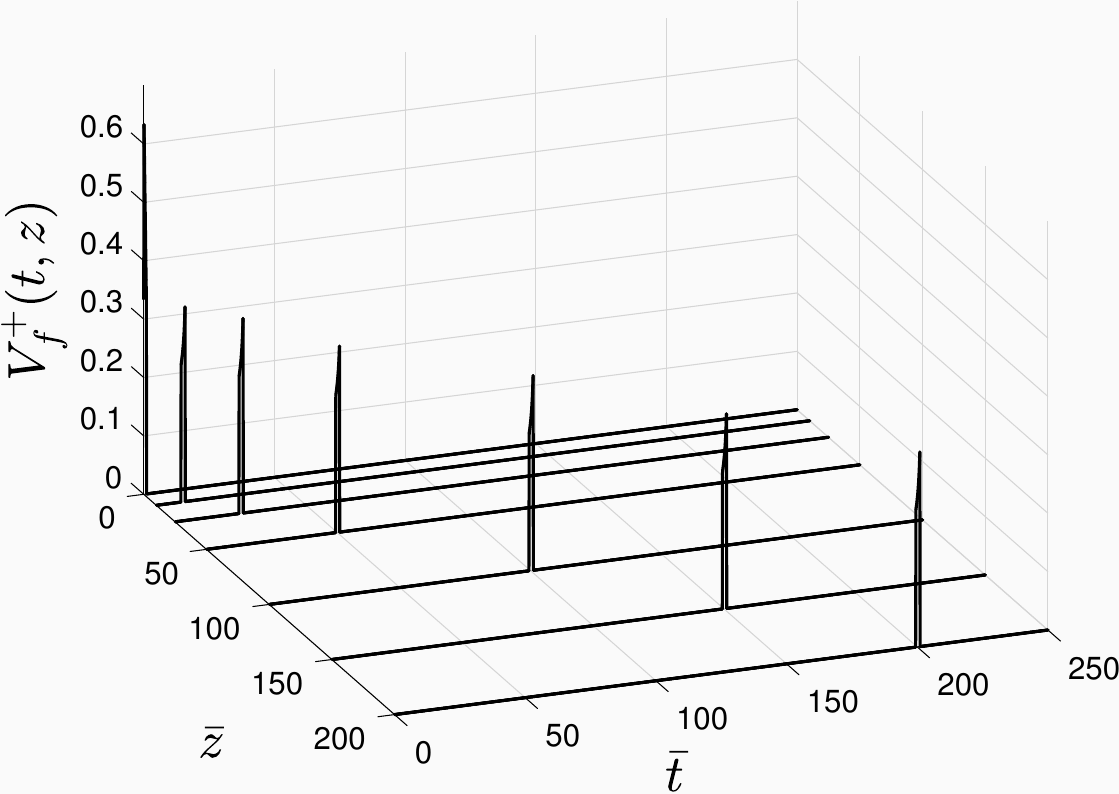}}
\hspace{5mm}
   \subfigure[$\zeta_0=5$]{%
     \label{Fig.Exmpl1.b}%
     \includegraphics[height=50mm]{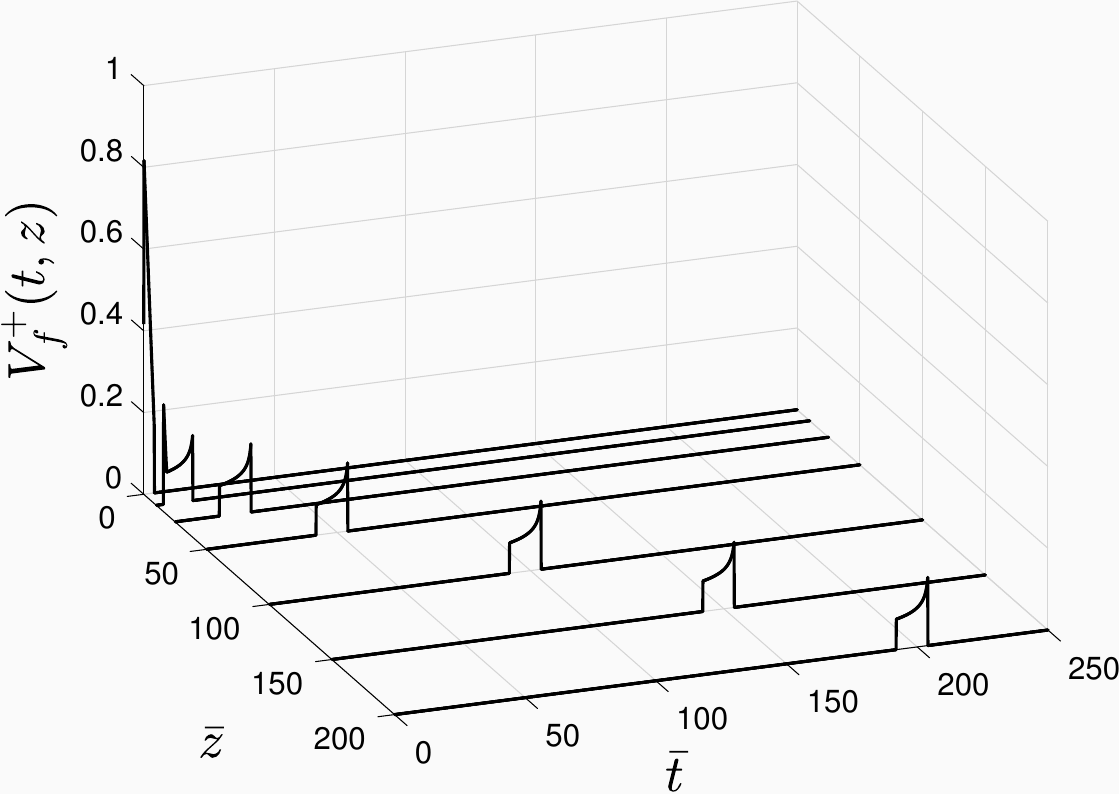}}
\vspace{1ex}
\centering
   \subfigure[$\zeta_0=10$]{%
     \label{Fig.Exmpl1.c}%
     \includegraphics[height=50mm]{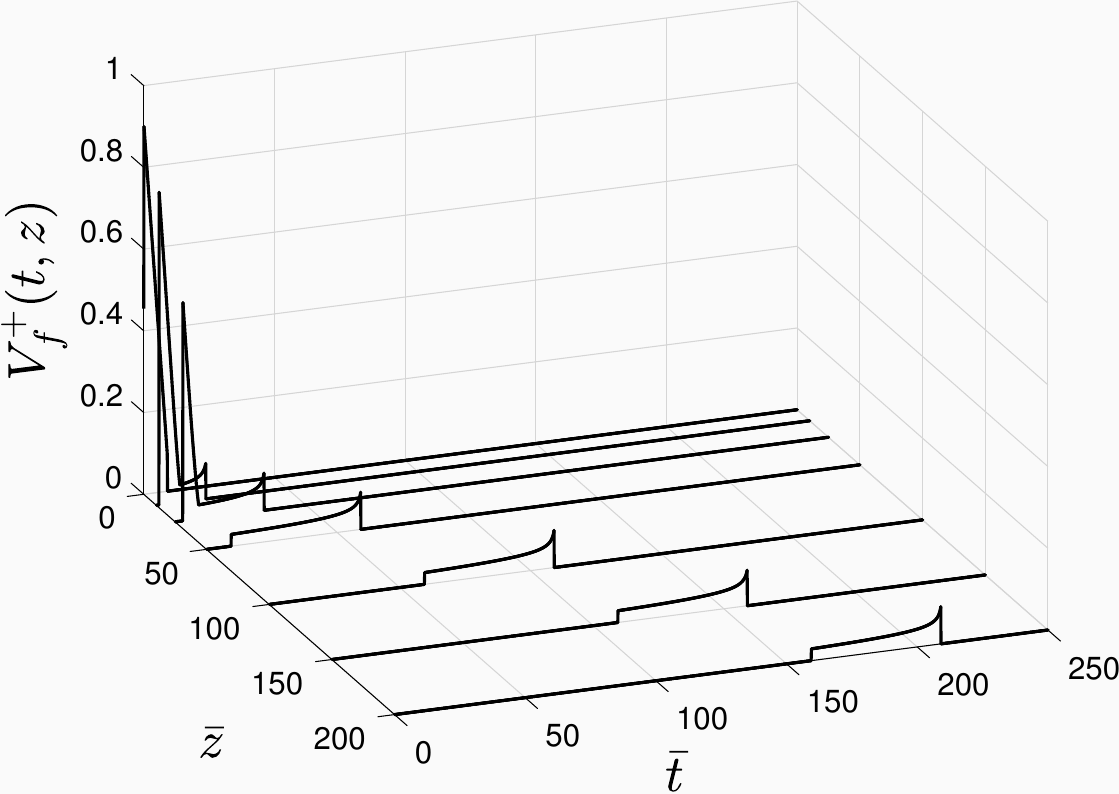}}
\hspace{5mm}
   \subfigure[$\zeta_0=20$]{%
     \label{Fig.Exmpl1.d}%
     \includegraphics[height=50mm]{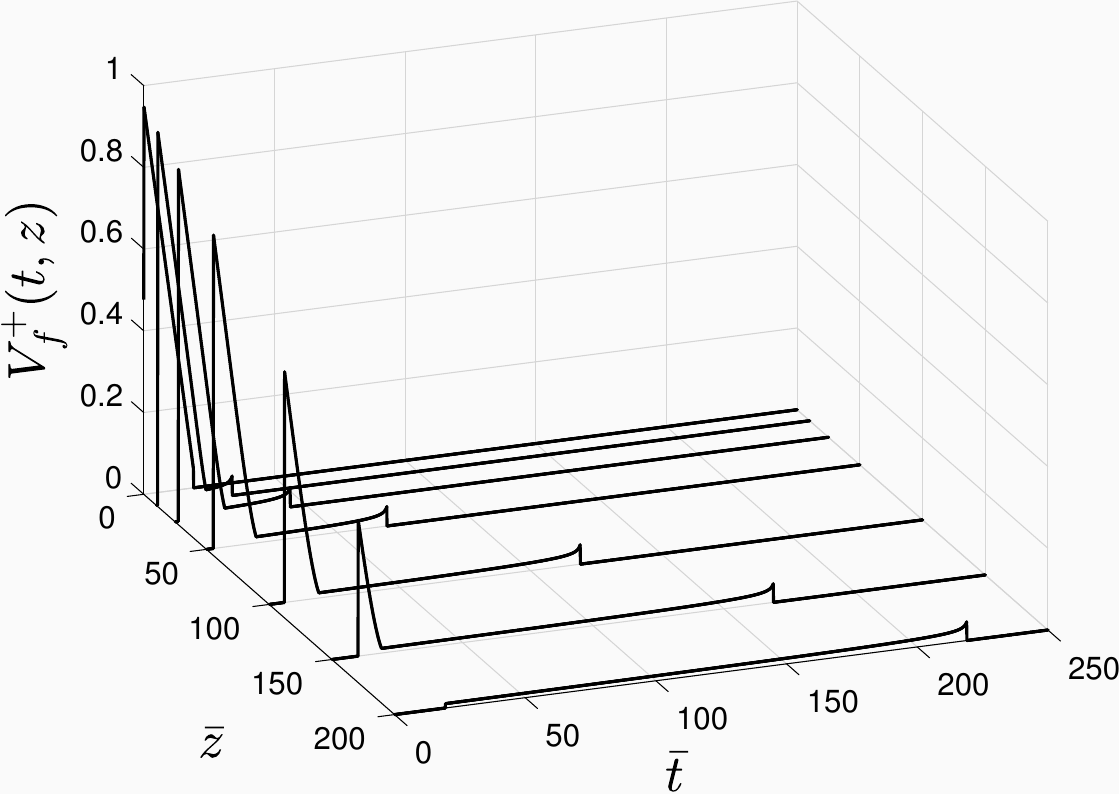}}
\caption[]{Normalized space-time snapshots of the pulse propagation in the softly time varying TL. The axes are normalized such that $\6z=z/l_a$ with $l_a=\tau_a v(t_m)=v_m\tau_a$ and $\6t=t/\tau_a$.}
\label{Fig.Exmpl1}
\end{figure*}

Figure~\ref{Fig.Exmpl1} shows space-time snapshots of the wave propagation in the capacitor's discharging circuit (Fig.~\ref{Fig.Discharge}) in this scenario. We set $Q_0/C_a=1$ and $\zeta_m=1$, and explore several cases: $\zeta_0=2,5,10,20$ in Figs.~\ref{Fig.Exmpl1.a}, \ref{Fig.Exmpl1.b}, \ref{Fig.Exmpl1.c}, \ref{Fig.Exmpl1.d}. As in Figure~\ref{Fig.Exmpl2}, the temporal axis $\6t=t/\tau_a$ is normalized with respect to $\tau_a$, but, the spatial axis $\6z=z/l_a=0,10,25,50,100,150,200$ is normalized with respect to $l_a=\tau_a v_m$ where $v_m=v(t_m)=(R_a/L_0)\zeta_m$. Though the switching scheme is different than that used in Fig.~\ref{Fig.Exmpl2}, similar effects can be noted. The difference between the figures comes due to $(i)$ different interplay between the $\zeta(t)$ and $\zeta(\tau_0(z,t))$ terms (compare \Eqref{eq.exmpl.7} and \Eqref{eq.exmpl.4}); $(ii)$ nonlinearity of the arrival time $\tau_0(z,t)$ following from \Eqref{eq.tau_0} due to the velocity variation during switching; and $(iii)$ the increase in the waveform width, i.e., ``pulse up scaling'', i.e., expansion due to the change in the TL wave-propagation velocity. Noting that the velocity changes linearly (smoothly) between $v_0=v(t_0)=(R_a/L_0)\zeta_0$ and $v_m=v(t_m)$, it can be shown that pulse width scaling ratio, i.e., the ratio of the pulse width after and before to switching is given by $(\zeta_0+\zeta_m)/2\zeta_m$, suggesting an increase of $1.5, 3, 5.5, 10.5$ for $\zeta_0=2,5,10,20$ in Figs.~\ref{Fig.Exmpl1.a}, \ref{Fig.Exmpl1.b}, \ref{Fig.Exmpl1.c}, \ref{Fig.Exmpl1.d}, respectively\footnote{This scaling ratio follows by observing that: $(i)$ modeling the soft switching as a series of abrupt switchings where at each such switches the pulse width changes by the ratio $v(t_s^-)/v(t_s^+)$, with $v(t_s^-)$ and $v(t_s^+)$ are the velocities before and after the switching time $t_s$, see in, \Eqref{eq.abrupt.a}; $(ii)$ recalling that such series of velocity changes in addition to a continuous (incremental) ``injection'' of signal at the TL's terminal due to the external source renders an infinitesimal temporal recursive-like increase in the propagating pulse width; and finally $(iii)$ approaching the continuity limit as was carried out in previous sections for the analysis of waveforms.}.

\vspace{1ex}
\noindent$\bullet$ \textbf{Inductor only switching scheme:} The dual, more difficult to implement, time varying inductor case with constant capacitance $C(t)=C_0$ gives inductance $L(t)=C_0 [Z(t)]^2$ and $v(t)=\left [C_0R_a \zeta(t) \right ]^{-1}$. Note that $v(t)$ is a monotonic increasing function of the time from an initial value $\left [C_0R_a \zeta_0 \right ]^{-1}$ to its final value $\left [C_0R_a \zeta_m \right ]^{-1}$. Similar to the discussion in the previous case, the corresponding wave-field is, thus, given by
\be
\begin{split}
&V_f^+(z,t)=\sqrt{ \frac{Z(\tau_0)}{Z(t)}} \, V_0(\tau_0)
\\[1ex]
&
= \frac{Q_0}{C_a(1+\zeta_0)}\, \frac{\bigl [ \zeta(\tau_0(z,t)) \bigl]^{3/2}}{\sqrt{\zeta(t)}}\times
\\[1ex]
&[ H(\tau_0(z,t)) -H(\tau_0(z,t)-t_m)\bigl].
\end{split}
\label{eq.exmpl.8}
\ee

Though the behavior of the impedance/velocity here is different than that treated in the previous case, corresponding switching effects, though different in details, can be discern, i.e., pulse compression, waveform deformation, etc. Thus, for brevity, they will not be further discussed here.

\begin{figure}[!h]
\centering
\includegraphics[width=80mm]{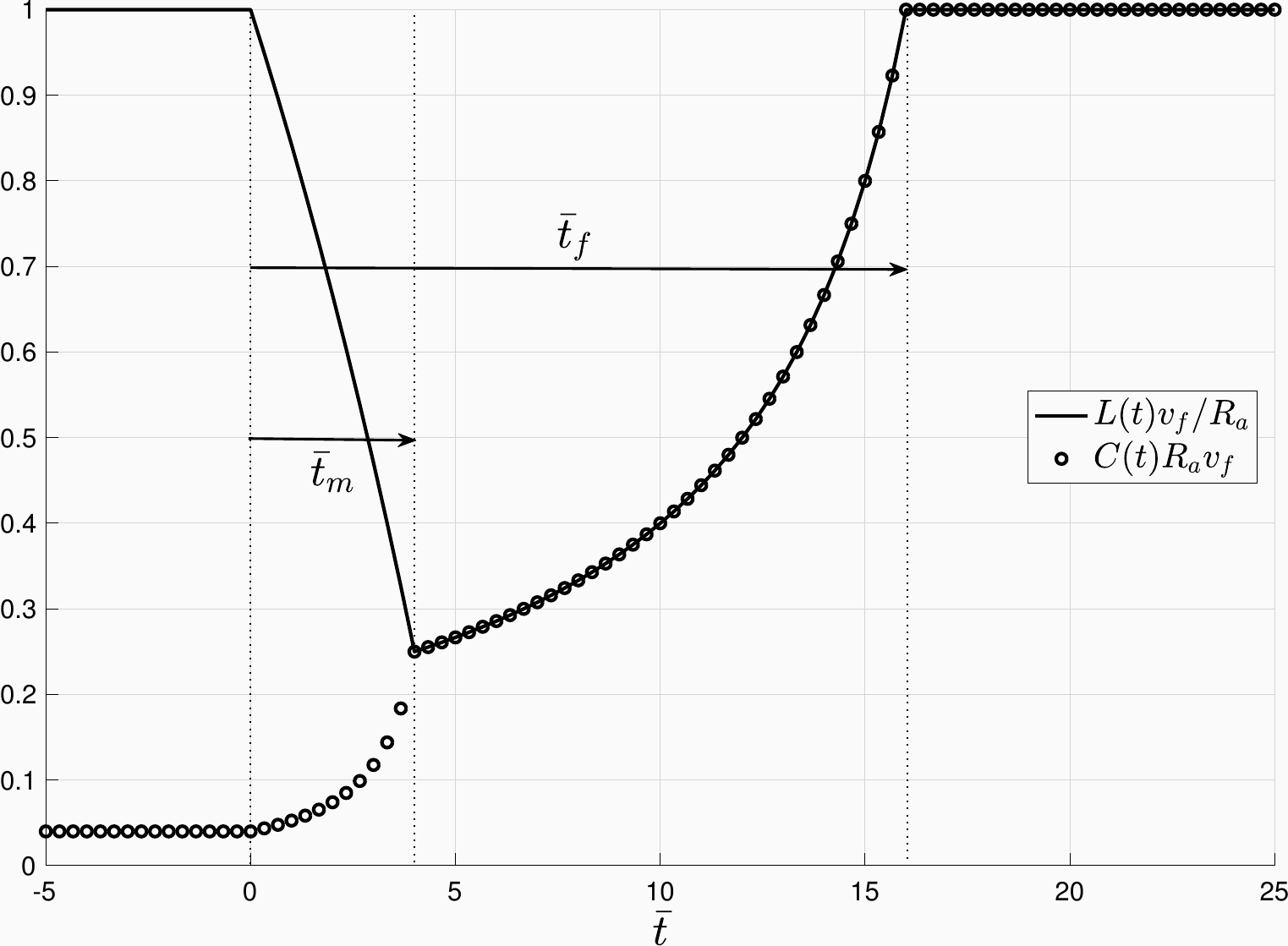}
\caption[]{Normalized characteristic inductance $L(t)$ and capacitance $C(t)$ per-unit-length of the switched TL for $\zeta_0=5$, $\zeta_m=1$ with $t_m=(\zeta_0-\zeta_m)\tau_a=4\tau_a$ and $v_0/v_f=5$ with $t_f=16 \tau_a$.}
\label{Fig.Exmpl3}
\end{figure}

\begin{figure*}[!ht]
\centering
   \subfigure[]{%
     \label{Fig.Exmpl4.a}%
     \includegraphics[height=60mm]{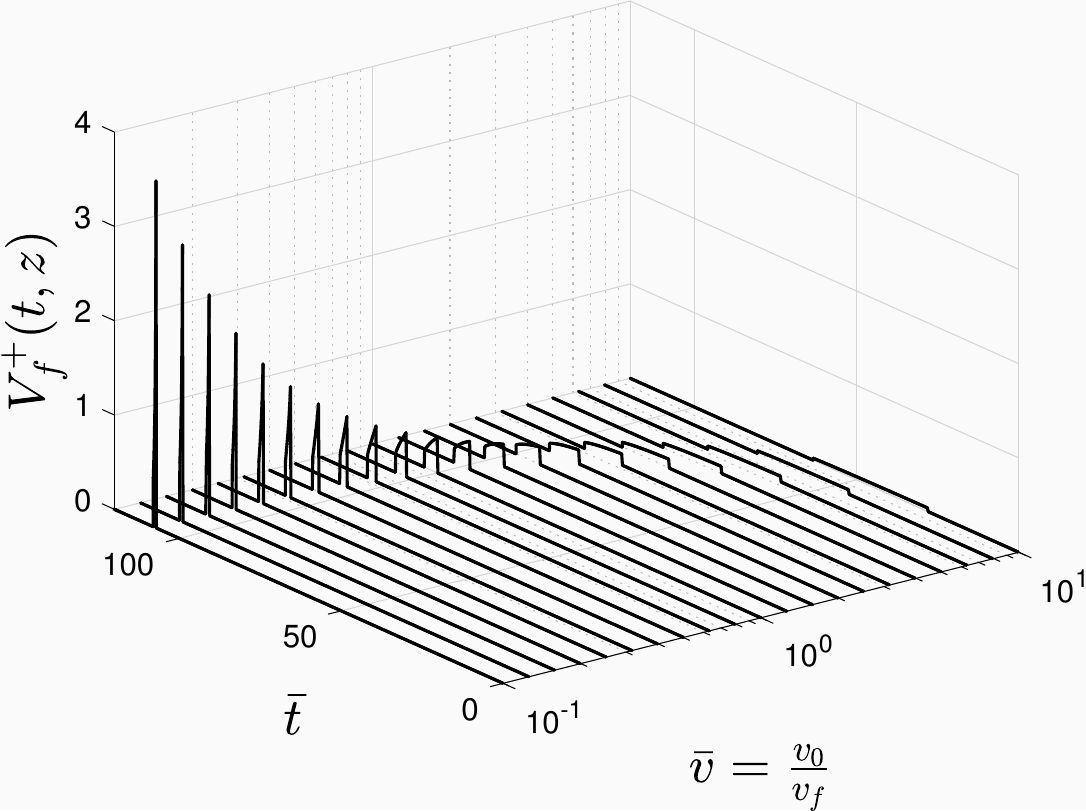}}
\hspace{5mm}
   \subfigure[]{%
     \label{Fig.Exmpl4.b}%
     \includegraphics[height=60mm]{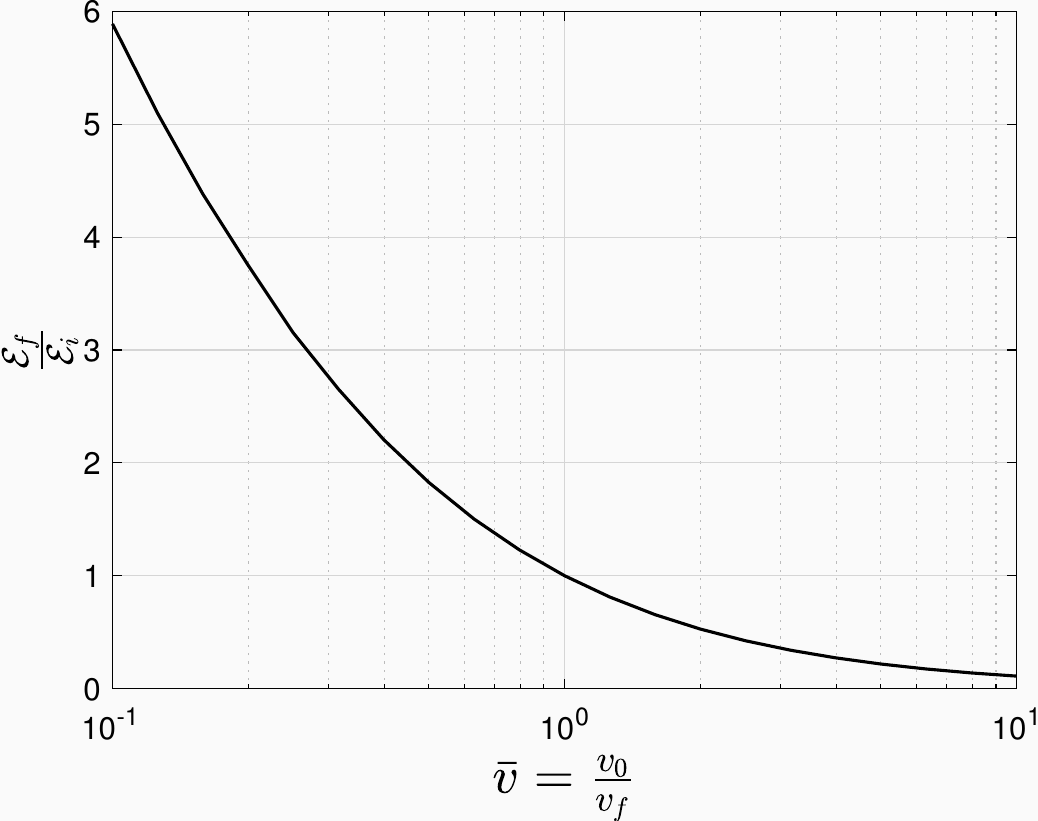}}
\caption[]{Waveforms for different ratios of initial to final velocities, $\6v=v_0/v_f$ in \ref{Fig.Exmpl4.a} and ratio of the energy of the forward propagating wave $\5E_f=\|V_f^+(z,t)\|^2/Z_m$ to the energy injected into the TL by the discharging capacitor $\5E_i=\|I_d \sqrt{Z(t)}\|^2$ in \ref{Fig.Exmpl4.a} for $\zeta_0=5$, $\zeta_m=1$ with $t_m=4\tau_a$ and $t_f=16 \tau_a$ and at observation point $z$ far from   the line termination.}
\label{Fig.Exmpl4}
\end{figure*}

\vspace{1ex}
\noindent$\bullet$ \textbf{Impedance and velocity change:} In this example we explore a case of non-coordinated temporal variation of the TL's velocity and characteristic impedance. The impedance changes {softly} from an initial, $Z_0$, to final, $Z_m$, values over a duration $t_m$ as described in \Eqref{eq.exmpl.3}. {The velocity, on the other hand,} changes from an initial value $v_0$ at time $t=0$ to a final value $v_f$ at time $t_f$ {(which is generally different than $t_m$)}, along a linear trajectory:
\be
\begin{split}
v(t)&=v_0 -(v_0-v_f)H(t-t_f)
\\[1ex]
&-(v_0-v_f) \frac{t}{t_f} \left[ H(t)-H(t-t_f)\right].
\end{split}
\label{eq.exmpl.9}
\ee
The velocity and impedance change is achieved by temporal gradual switching of the TL's capacitance and inductance per-unit-length as depicted in Fig.~\ref{Fig.Exmpl3} for the case of $\zeta_0=5$, $\zeta_m=1$ with $t_m=(\zeta_0-\zeta_m)\tau_a=4\tau_a$ and $v_0/v_f=5$ with $t_f=16 \tau_a$. The nonlinear trajectory of $L(t)$ and $C(t)$ is due to the different rates and durations of changes of the impedance and velocity.   Figure~\ref{Fig.Exmpl4.a} depicts waveforms of the forward propagating wave for different ratios of initial to final velocities, $\6v=v_0/v_f$, for $\zeta_0=5$, $\zeta_m=1$ with $t_m=4\tau_a$ and $t_f=16 \tau_a$ and at observation point $z$ such that the arrival of the wavefront takes place  after the switching ends. Figure~\ref{Fig.Exmpl4.b} depicts the ratio of the energy of the forward propagating wave $\5E_f=\|V_f^+(z,t)\|^2/Z_m$ to the energy injected into the TL by the discharging capacitor $\5E_i=\|I_d \sqrt{Z(t)}\|^2$ where $\| \ \|$ is the energy norm as in \Eqref{eq.abrupt.c}. Following the discussion in the previous examples, it is interesting to note in Fig.~\ref{Fig.Exmpl4} that whenever $v_0 > v_f$ energy is absorbed from the wave system and the pulse width widened whereas for $v_0 < v_f$ energy is injected into the wave system by the switching of the TL's properties and the pulse width become narrower. Whenever no change in the velocity occurs during switching , i.e., $v_f=v_0$, no energy is injected nor absorbed by the switching system ($\5E_f/\5E_i=1$). This behaviour of no energy exchange with the switching system is in accordance with \Eqref{eq.enrg.blnc.2}. Moreover, this case corresponds also to the discussion above in the first example of ``Constant velocity scheme''.

%
To conclude this example section, it should be emphasized that the analysis here assumes that any back-reflected wave due to the forward propagation wave in this time-varying media is negligibly small and does not cause major deformations to the forward, first order, WKB contributions.

\section{Discussion and Conclusions}
\label{sec.DISCON}
In this work we have developed analytical formulas describing the transmitted and reflected waves due to a propagating short-pulse that experiences gradual temporal switching of the guiding medium. The analysis is based on the WKB approach, developed originally as a general tool for differential equations with softly varying coefficients. While this approach has been extensively used in quantum-mechanics and for electromagnetic and acoustic wave propagation in spatially varying media, its application to time-varying wave systems was limited so far.
We have developed the WKB formalism for medium that experiences simultaneous and independent variation of both impedance and wave velocity (capacitance and inductance per-unit-length or, synonymously, permittivity and permeability). We have moreover extended our formalism to be applicable also for excitation problems by a lumped source, using the concept of impulse response and superposition in linear time-varying systems.
Finally, we concluded by demonstration of the analytical formalism on a canonical problem of capacitor discharge into a semi-infinite transmission line. These examples demonstrate that gradual change of the velocity enables energy absorption (or harvesting) from the wave system to the switching system (with negligible reflected wave-field) and can form the basis for an energy based electromagnetic absorber/shield of pulsed energy. The opposite case where energy is invested into the wave system by the switching system implies a type of parametric amplification or an energy based wave accelerator. Both cases have numerous application in electromagnetics as well as in acoustics.

\section*{Appendix}
\label{sec.APPENDIX}
This appendix describes the derivation of the discharge current $I_d$ and the TL's characteristic impedance $Z(t)$ in \Eqref{eq.exmpl.1}. In the present discussion the interest is only on the capacitor's discharge, thus the TL can be modeled by its characteristic impedance and to appear as a lumped impedance element $Z(t)$. The design goal in the discharge process is to obtain an optimal energy delivery from an initially charged capacitor to the TL, i.e., $Z(t)$.

The capacitor's charge $Q(t)$ during discharge is given by
\be
Q(t)=-Q_0 e^{-\int_0^t dt' [\tau(t')]^{-1}}, \quad \tau_t=C_a(R_a+Z(t)).
\label{eq.app.1}
\ee
where we assumed that initially at $t=0$ the capacitor is charged with a negative charge $-Q_0$. The discharge current is $i(t)=dQ(t)/dt=-Q(t)/\tau(t)$, the instantaneous power delivered to the TL is $p(t)=i^2(t)Z(t)=Q^2(t) Z(t)/\tau^2(t)$ and the energy delivered during discharge is
\begin{eqnarray}
\mathcal{E}&=&\int_0^\infty dt \left (\frac{Q(t)}{\tau(t)} \right )^2 Z(t) \nonumber\\
&=& \frac{R_a Q_0^2}{\tau_a} \int_0^\infty dt \left [ y'(t)-\tau_a(y'(t))^2\right ]e^{-2y(t)}.
\label{eq.app.2}
\end{eqnarray}
where $y(t)=\int_0^t dt' [\tau(t')]^{-1}$ and $Q(t)=-Q_0 e^{-y(t)}$.

To obtain an optimal solution to the capacitor discharge and energy delivery to $Z(t)$, the corresponding Euler-Lagrange equation \cite{Gelfand1963} is formulated:
\be
\begin{split}
&\frac{d}{dy} \mathcal L(y,y',t)-\frac{d}{dt} \frac{d}{dy'}\mathcal L(y,y',t) =0,
\\[1ex]
&\mathcal L(y,y',t) = \left [ y'(t)-\tau_a(y'(t))^2\right ]e^{-2y(t)},
\end{split}
\label{eq.app.3}
\ee
with $y'(t)=\frac{d y(t)}{dt}$. Using $\mathcal L(y,y',t)$ in Euler-Lagrange equation gives the differential equation
\be
y''(t)-(y'(t))^2=0.
\label{eq.app.4}
\ee
Noting  that $\tau(t)=\tau_a(1+\zeta(t))=(y'(t))^{-1}$ with $\zeta(t)=Z(t)/R_a$, see, e.g., \Eqref{eq.exmpl.1} gives
\be
\zeta(t)=\frac{1}{\tau_a y'(t)}-1.
\label{eq.app.5}
\ee
Assuming that initially at $t=0$ the transmission line characteristic impedance is $Z_0=R_a \zeta_0$, it gives the initial condition for the solution of the differential equation in \Eqref{eq.app.4}: $y'(0)=[\tau_a(1+\zeta_0)]^{-1}$. It further follows that
\bse
\label{eq.app.6}
\bea
&y'(t)=\frac{1}{\tau_a(1+\zeta_0)-t},
\label{eq.app.6a}
\\[2ex]
&y(t)=-\ln \left [ 1- \frac{t}{\tau_a(1+\zeta_0)}\right ],
\label{eq.app.6b}
\eea
\ese
and in view of \Eqref{eq.app.5} it gives $\zeta(t)$ of \Eqref{eq.exmpl.1}. The capacitor charge is, thus  given by
\be
Q(t)=-Q_0 \left [ 1-\frac{t}{\tau_a(1+\zeta_0)}\right ],
\label{eq.app.7}
\ee
leading to the constant discharge current $I_d=\frac{Q_0}{\tau_a(1+\zeta_0)}$. It follows that a complete discharge of the capacitor is achieved at $t_d=\tau_a(1+\zeta_0)$. However, at such $t_d$, $\zeta(t_d)<0$ ($Z(t_d)<0$) which is not physical. Furthermore, achieving $\zeta \to 0$ is impractical, hence we set the lowest possible characteristic impedance as $Z_m=R_a \zeta_m >0$ which is reached at time $t_m=(\zeta_0-\zeta_m)\tau_a$.


\begin{thebibliography}{00}


\bibitem{Collin}
R. E. Collin, \emph{Foundations for Microwave Engineering}, McGraw-Hill, 1966.

\bibitem{Kinsler}
L. E. Kinsler, and A. R. Frey, \emph{Fundamentals of Acoustics}, John Wiley \& Sons, 2nd ed. 1962.



\bibitem{Morgenthaler1958}
F. R. Morgenthaler, ``Velocity modulation of electromagnetic waves,'' \emph{IRE Trans. Microw. Theory Tech.}, {\bf 6}, pp. 167-172 (1958).


\bibitem{Weinstein1965}
H. Weinstein, ``Linear signal stretching in a time-variant system,'' \emph{IEEE Trans. Circ. Theory} {\bf12} pp. 157-164 (1965).

\bibitem{Auld1968}
B. A. Auld, J. H. Collins, and H. R. Zapp, ``Signal processing in a non-periodically time-varying magnetoelastic medium,'' \emph{Proc. IEEE} {\bf 56}, pp. 258-272 (1968).



\bibitem{Rezende1969}
S. M. Rezende and F. R. Morgenthaler, ``Magnetoelastic waves in time-varying magnetic field I: Theory,'' \emph{J. Appl. Phys.} {\bf 40}, pp. 524-536 (1969).



\bibitem{Felsen1970}
L. B. Felsen, G. M. Whitman, ``Wave propagation in time-varying media,'' \emph{IEEE Trans. Ant. Prop.} {\bf 18}, 2, pp. 242-253 (1970).


\bibitem{Fante1971}
R. L. Fante, ``Transmission of electromagnetic Waves into time-varying media,'' \emph{IEEE Trans. Ant. Prop.}, {\bf 19}, No. 3 pp. 417-424 (1971).

\bibitem{Agrawal2014}
Y. Xiao, D. N. Maywar, and G. P. Agrawal, ``Reflection and transmission of electromagnetic waves at a temporal boundary,'' \emph{Opt. Lett.}, {\bf39}, No. 3 (2014).

\bibitem{Budko2009}
Neil V. Budko, ``Electromagnetic radiation in a time-varying background medium,'' \emph{Phys. Rev. A}, {\bf80}, 053817 (2009).


\bibitem{Cervantes-Gonzalez2009}
J. R. Zurita-Sanchez, P. Halevi, and J. C. Cervantes-Gonzalez, ``Reflection and transmission of a wave incident on a slab with a time-periodic dielectric function $\epsilon(t)$,'' \emph{Phys. Rev. A}, {\bf79}, 053821 (2009).


\bibitem{Kunz1966}
D. E. Holberg, and K. S. Kunz, ``Parametric properties of fields in a slab of time-varying permittivity,'' \emph{IEEE Trans. Ant. Prop.}, {\bf14}, No. 2, pp. 183-194 (1966).

\bibitem{Chegnizadeh2018}
M. Chegnizadeh, K. Mehrany, and M. Memarian, ``General solution to wave propagation in media undergoing arbitrary transient or periodic temporal variations of permittivity,'' \emph{J. Opt. Soc. Am. B} {\bf 35}, pp. 2923-2932 (2018)

\bibitem{Hayrapetyan2016}
A. G. Hayrapetyan, J. B. G{\"o}tte, K. K. Grigoryan, S. Fritzsche, and R. G. Petrosyan, ``Electromagnetic wave propagation in spatially homogeneous yet smoothly time-varying dielectric media,'' \emph{Journal of Quantitative Spectroscopy and Radiative Transfer}, {\bf 178}, pp. 158-166, (2016)

\bibitem{Koutserimpas2018}
T. T. {Koutserimpas} and R. {Fleury}, ``Electromagnetic Waves in a Time Periodic Medium With Step-Varying Refractive Index,'' \emph{IEEE Trans. Ant. Prop.,} {\bf 66}, (10), pp. 5300-5307, (2018).

\bibitem{Lurie2016}
Konstantin A. Lurie and Vadim V. Yakovlev, ``Energy accumulation in waves propagating in space- and time-varying transmission lines,'' \emph{IEEE Ant. Wireless Prop. Lett.,} {\bf15} pp. 1681-1684 (2016).

\bibitem{Lurie2017}
K. A. Lurie, D. Onofrei, W. C. Sanguinet, S. L. Weekes, and Vadim V. Yakovlev, ``Energy accumulation in a functionally graded spatial-temporal checkerboard,'' \emph{IEEE Ant. Wireless Prop. Lett.,}{\bf16} pp. 1496-1499 (2017).

\bibitem{Tretyakov2018}
M. S. Mirmoosa, G. A. Ptitcyn, V. S. Asadchy and S. A. Tretyakov, ``Unlimited accumulation of electromagnetic energy using time-varying reactive elements,'' \emph{arXiv:1802.07719v1} (2018).


\bibitem{Fleury2018b}
T. T. Koutserimpas, and R. Fleury, ``Electromagnetic waves in a time periodic medium With step-varying refractive index,'' \emph{IEEE Trans. Ant. Prop.,} {\bf 66} (10), (2018).

\bibitem{Engheta2018Cleo}
Y. Kiasat, V. Pacheco-Pena, B. Edwards, and N. Engheta, ``Temporal metamaterials with non-Foster networks,'' in Conference on Lasers and Electro-Optics, OSA Technical Digest (online) (Optical Society of America, 2018), paper JW2A.90.


\bibitem{Fleury2018}
T. T. Koutserimpas and R. Fleury, ``Nonreciprocal gain in non-Hermitian time-Floquet systems,'' \emph{Phys. Rev. Lett.}, {\bf120}, 087401 (2018).



\bibitem{OL_Caloz2018}
A. Akbarzadeh, N. Chamanara, and C. Caloz, ``Inverse prism based on temporal discontinuity and spatial dispersion,'' \emph{Opt. Lett.,} {\bf 43}, 14, pp. 3297-3300 (2018).

\bibitem{Chamanara2019}
N. {Chamanara} and Y. {Vahabzadeh} and C. {Caloz}, ``Simultaneous Control of the Spatial and Temporal Spectra of Light With Space-Time Varying Metasurfaces,'' \emph{IEEE Trans. Ant. Prop.,} {\bf 67} (4), pp. 2430-2441 (2019).


\bibitem{Halevi2016}
Juan Sabino Martınez-Romero, O. M. Becerra-Fuentes, and P. Halevi, ``Temporal photonic crystals with modulations of both permittivity and permeability,'' \emph{Phys. Rev. A}, {\bf93}, 063813 (2016).


\bibitem{Skorobogatiy2016}
H. Qu, Zo\'{e}-L. Deck-L\'{e}ger, C. Caloz, and M. Skorobogatiy, ``Frequency generation in moving photonic crystals,'' \emph{J. Opt. Soc. Am. B,} {\bf 33}, pp. 1616-1626 (2016).

\bibitem{PRACaloz2018}
N. Chamanara, Zo\'{e}-L. Deck-L\'{e}ger, C. Caloz, and D. Kalluri, ``Unusual electromagnetic modes in space-time-modulated dispersion-engineered media,'' \emph{Phys. Rev. A,} {\bf 97}, 063829 (2018).

\bibitem{Caloz2017}
S. Taravati and C. Caloz, ``Mixer-duplexer-antenna leaky-wave system based on periodic space-time modulation,'' \emph{IEEE Tran. Ant. Prop.}, {\bf 65}, No. 2, pp. 442-452 (2017).



\bibitem{Fort2016}
V. Bacot, M. Labousse, A. Eddi, M. Fink, and E. Fort, ``Time reversal and holography with
spacetime transformations,'' \emph{Nat. Phys.}, {\bf12}, pp. 972-977 (2016).


\bibitem{Mattei2017}
G. W. Milton, and  O. Mattei, ``Field patterns: a new mathematical object,'' \emph{Proc. R. Soc. A} {\bf473}:20160819 (2017).


\bibitem{Fang2012}
K. Fang, Z. Yu, S. Fan, ``Realizing effective magnetic field for photons by controlling the phase of dynamic modulation,'' \emph{Nat. phot.} {\bf6} (11), 782 (2012).



\bibitem{Sounas2013}
D. L. Sounas, C. Caloz, A. Al\'{u}, ``Giant non-reciprocity at the subwavelength scale using angular momentum-biased metamaterials,'' \emph{Nat. comm.} {\bf4}, 2407 (2013).


\bibitem{Fleury2014}
R. Fleury, D. L. Sounas, C. F. Sieck, M. R. Haberman, A. Al\'{u}, ``Sound isolation and giant linear nonreciprocity in a compact acoustic circulator,'' \emph{Science} {\bf343} (6170), pp. 516-519	(2014).

\bibitem{Estep2014}
N. A. Estep, D. L. Sounas, J. C. Soric, A. Al\'{u}, ``Magnetic-free non-reciprocity and isolation based on parametrically modulated coupled-resonator loops,'' \emph{Nat. Phys.} {\bf10} (12), pp. 923-926	(2014).

\bibitem{Hadad2015}
Y. Hadad, D. L. Sounas, and A. Al\'{u}, ``Space-Time Gradient Metasurfaces,'' \emph{Phys. Rev. B} {\bf 92} (10), 100304(R) (2015).


\bibitem{Hadad2016}
Y. Hadad, J. C. Soric, A. Al\'{u}, ``Breaking temporal symmetries for emission and absorption,''
\emph{Proceedings of the National Academy of Sciences} {\bf113} (13), 3471-3475 (2016).


\bibitem{Taravati2017}
S. Taravati, N. Chamanara, and C. Caloz, ``Nonreciprocal electromagnetic scattering from a periodically space-time modulated slab and application to a quasisonic isolator,'' \emph{Phys. Rev. B} {\bf 96}, (16) (2017).



\bibitem{Sounas2017}
D. L. Sounas, A. Al\'{u}, ``Non-reciprocal photonics based on time modulation,'' \emph{Nat. Phot.} {\bf11} (12), 774-783 (2017).


\bibitem{Nagulu2018}
A. Nagulu , T. Dinc , Z. Xiao, M. Tymchenko, D. L. Sounas, A. Alu, and H. Krishnaswamy, ``Non-reciprocal components based on switched transmission lines,'' \emph{IEEE Trans. Microw. Theory Tech.}, (2018).

\bibitem{Fan2018a}
S. Fan, Y. Shi, Q. Lin, ``Nonreciprocal photonics without magneto-optics,'' \emph{IEEE Ant. Wireless Prop. Lett.}, to appear, 2018


\bibitem{Fan2018b}
M. Minkov and S. Fan, ``Localization, time-reversal, and unidirectional guiding of light pulses using dynamic modulation,'' in Conference on Lasers and Electro-Optics, OSA Terchnical Digest (online) (Optical Society of America, 2018), paper FM1E.4.

\bibitem{Fan2018c}
M. Minkov and S. Fan, ``Localization and time-reversal of light through dynamic modulation,''
\emph{Phys. Rev. B,} {\bf97}, 060301(R),  (2018).



\bibitem{Shlivinski2018}
A. Shlivinski and Y. Hadad, ``Beyond the Bode-Fano bound: Wideband impedance matching for short-pulses using temporal switching of transmission-line parameters,'' \emph{Phys. Rev. Lett}, {\bf 121}, 204301, (2018).


\bibitem{Bode}
H. W. Bode, \emph{Network analysis and feedback amplifier design}, Van Nostrand, New York, 1945.


\bibitem{Fano}
R. M. Fano, ``Theoretical limitations on the broadband matching of arbitrary impedances'', Technical Report NO. 41, Research Laboratory of Electronics, MIT, 1948.


\bibitem{Acher2009}
O. Acher, J. M. L. Bernard, P. Maréchal, A. Bardaine, F. Levassorta, ``Fundamental constraints on the
performance of broadband ultrasonic matching structures and absorbers,''
\emph{The Journal of the Acoustical Society of America} {\bf125}, pp.~1995-2005 (2009).

\bibitem{Pozar}
D. M. Pozar, \emph{Microwave Engineering}, John Wiely \& Sons, 2nd ed. 1998 (p. 295-297),



\bibitem{PaiYen}
P.-Y. Chen, J. Soric, A. Al\'{u}, ``Invisibility and cloaking based on scattering cancellation,'' \emph{Advanced Materials} {\bf24} (44) pp.281-304 (2012).

\bibitem{Fleury2014_PRB}
R. Fleury, J. Soric, A. Alu, ``Physical bounds on absorption and scattering for cloaked sensors'',\emph{Phys. Rev. B.,} {\bf 89}, 045122 (2014).

\bibitem{Monticone}
F. Monticone, A. Al\'{u}, ``Invisibility exposed: physical bounds on passive cloaking'',
\emph{Optica} {\bf3} (7), 718-724 (2016).



\bibitem{Chu}
L. J. Chu, "Physical limitations of omni-directional antennas," \emph{J. App. Phys.}, {\bf 19}, 12, pp. 1163–1175, (1948).

\bibitem{Harrington}
R. F. Harrington, ``Effect of antenna size on gain, bandwidth, and efficiency,'' \emph{J. Res. Nat. Bur. Stand}, {\bf 64}, 1, pp. 1–12, (1960).

\bibitem{Gustafsson2007}
M. Gustafsson, C. Sohl, G. Kristensson, ``Physical limitations on antennas of arbitrary shape,'' \emph{Proceedings of the Royal Society of London A: Mathematical, Physical and Engineering Sciences}, {\bf 463}, 2086, pp. 2589-2607 (2007).


\bibitem{Gustafsson2009}
M. Gustafsson, C. Sohl,  and G. Kristensson, ``Illustrations of new physical bounds on linearly polarized antennas,'' \emph{IEEE Trans. Ant. Prop.,} {\bf 57}, 5, (2009).

\bibitem{Gustafsson2015}
M. Gustafsson, Tayli D., and M. Cismasu, \emph{Physical Bounds of Antennas}, Handbook of Antenna Technologies. Editor Zhi N. Chen, Springer, (2015). pp. 1-32.


\bibitem{Thal2006}
H.L. Thal, ``New radiation Q-limits for spherical wire antennas'', \emph{IEEE Trans. Ant. Prop.}, {\bf 54}, pp. 2757-2763, October 2006.


\bibitem{Schab2018}
K. Schab, L. Jelinek, M. Capek, C. Ehrenborg, D. Tayli, Guy A. E.  Vandenbosch, ``Energy stored by radiating systems,'' \emph{IEEE Access}, {\bf 6}, pp. 10553-10568, (2018).






\bibitem{Capek2017}
M. Capek, M. Gustafsson and K. Schab, ``Minimization of antenna quality factor,'' \emph{IEEE Trans. Ant. Prop.}, {\bf65}, 8, pp. 4115-4123, (2017).




\bibitem{Wang2010}
X. J. Xu and Y. E. Wang, ``A direct antenna modulation (DAM) transmitter with a switched electrically small antenna,'' \emph{2010 International Workshop on Antenna Technology (iWAT),} Lisbon, 2010, pp.1-4.


\bibitem{Wang2014}
U. Azad, Y. E. Wang, ``Direct antenna modulation (DAM) for enhanced capacity performance of near-field communication (NFC) link,'' \emph{IEEE Trans. Circ. Sys.}, {\bf 61} (3), (2014).


\bibitem{Wang2017}
Srinivas Prasad M. N., Y. Huang,Y. Ethan Wang,  ``Going beyond Chu Harrington limit: ULF radiation with a spinning magnet array,'' \emph{32nd URSI GASS}, Montreal, 19–26 August 2017.



\bibitem{Shlivinski1997}
A. Shlivinski, E. Heyman, and R. Kastner, ``Antenna characterization in the time domain,'' \emph{IEEE Trans. Ant. Prop.}, {\bf45}, pp. 1140-1149, (1997).


\bibitem{Shlivinski1999a}
A. Shlivinski and E. Heyman, ``Time-domain near-field analysis of short-pulse antennas-part I: spherical wave (multipole) expansion,'' \emph{IEEE Trans. Ant. Prop.}, {\bf47}, 2, pp. 271-279, (1999).


\bibitem{Shlivinski1999b}
A. Shlivinski and E. Heyman, ``Time-domain near-field analysis of short-pulse antennas-part II: reactive energy and the antenna Q'', \emph{IEEE Trans. Ant. Prop.}, {\bf47}, 2, pp. 280-286, (1999)





\bibitem{Sjoberg2010}
M. Gustafsson, D. Sjoberg, ``Sum rules and physical bounds on passive metamaterials,'' \emph{New Journal of Physics}, {\bf 12} (4), 043046, (2010).

\bibitem{Sjoberg2011}
M. Gustafsson, D. Sjoberg, ``Physical bounds and sum rules for high-impedance surfaces,'' \emph{IEEE Trans. Ant. Prop.,} {\bf59}, (6), pp. 2196-2204, (2011).



\bibitem{Bremmer1951}
H. Bremmer, ``The W.K.B. approximation as the first term of a geometric-optical series'', \emph{Comm. Pure Appl. Math.}, {\bf4}, pp. 105-115, (1951).

\bibitem{Keller1962}
H. B. Keller and J. B. Keller, ``Exponential-Like Solutions of Systems of Linear Ordinary Differential Equations,'' \emph{J. Soc. Indust. Appl. Math.}, {\bf10}, (2), pp. 246-259 (1962).



\bibitem{Gelfand1963}
I.M. Gelfand, and S.V. Fomin, and R.A. Silverman, \emph{Calculus of Variations}, Prentice-Hall, Inc, 1963

%
%
%




%

%
%
%






%
%
%
%




%
%
%
%
%


%
%
%

%
%
%




%
%
%
%
%
%



























%





\end{thebibliography}
\end{document}